\documentclass[pdflatex,sn-mathphys-num]{sn-jnl}

\usepackage{graphicx}%
\usepackage{multirow}%
\usepackage{amsmath,amssymb,amsfonts}%
\usepackage{amsthm}%
\usepackage{mathrsfs}%
\usepackage[title]{appendix}%
\usepackage{xcolor}%
\usepackage{textcomp}%
\usepackage{manyfoot}%
\usepackage{booktabs}%
\usepackage{algorithm}%
\usepackage{algorithmicx}%
\usepackage{algpseudocode}%
\usepackage{listings}%

\usepackage{todonotes}
\usepackage[version=4]{mhchem}
\usepackage{subfigure}
\usepackage{orcidlink}

\theoremstyle{thmstyleone}%
\theoremstyle{thmstyletwo}%

\theoremstyle{thmstylethree}%
\begin{document}

\title[Article Title]{Numerical Investigation of Preferential Flow Paths in Enzymatically Induced Calcite Precipitation supported by Bayesian Model Analysis}


\author*[1]{\fnm{Rebecca} \sur{Kohlhaas}\orcidlink{0000-0001-5263-5073}}\email{rebecca.kohlhaas@iws.uni-stuttgart.de}

\author[1]{\fnm{Johannes} \sur{Hommel}\orcidlink{0000-0002-2061-0303}}\email{johannes.hommel@iws.uni-stuttgart.de}

\author[1,2]{\fnm{Felix} \sur{Weinhardt}\orcidlink{0000-0002-1758-9427}}\email{felix.weinhardt@ufz.de}

\author[1]{\fnm{Holger} \sur{Class}\orcidlink{0000-0002-4476-8017}}\email{holger.class@iws.uni-stuttgart.de}

\author[3]{\fnm{Sergey} \sur{Oladyshkin}\orcidlink{0000-0003-4676-5685}}\email{sergey.oladyshkin@iws.uni-stuttgart.de}

\author[1]{\fnm{Bernd} \sur{Flemisch}\orcidlink{0000-0001-8188-620X}}\email{bernd.flemisch@iws.uni-stuttgart.de}

\affil*[1]{\orgdiv{Institute for Modelling Hydraulic and Environmental Systems,  LH2}, \orgname{University of Stuttgart}, \orgaddress{\city{Stuttgart}, \country{Germany}}}

\affil[2]{\orgdiv{Department Technical Biogeochemistry}, \orgname{Helmholtz Centre for Environmental Research GmbH - UFZ}, \orgaddress{\city{Leipzig}, \country{Germany}}}

\affil[3]{\orgdiv{Institute for Modelling Hydraulic and Environmental Systems,  LS3}, \orgname{University of Stuttgart}, \orgaddress{\city{Stuttgart}, \country{Germany}}}

\abstract{The usability of enzymatically induced calcium carbonate precipitation (EICP) as a method for altering porous-media properties, soil stabilization, or biocementation depends on our ability to predict the spatial distribution of the precipitated calcium carbonate in porous media.
While current REV-scale models are able to reproduce the main features of laboratory experiments, they neglect effects like the formation of preferential flow paths and the appearance of multiple polymorphs of calcium carbonate with differing properties.
We show that extending an existing EICP model by the conceptual assumption of a mobile precipitate, amorphous calcium carbonate (ACC), allows for the formation of preferential flow paths when the initial porosity is heterogeneous.
We apply sensitivity analysis and Bayesian inference to gain an understanding of the influence of characteristic parameters of ACC that are uncertain or unknown and compare two variations of the model based on different formulations of the ACC detachment term to analyse the plausibility of our hypothesis.
An arbitrary Polynomial Chaos (aPC) surrogate model is trained based on the full model and used to reduce the computational cost of this study. 
\paragraph{Article highlights}
\begin{enumerate}
    \item REV-scale EICP model that was extended with mobile ACC is able to produce preferential flow paths.
    \item Initial heterogeneity is required for the formation of preferential flow paths.
    \item Exploration of dependency on unknown model parameters by surrogate-assisted Bayesian model analysis.
\end{enumerate}
}




\maketitle

\section{Introduction}\label{sec:intro}
Biomineralization processes such as Enzymatically Induced Calcium carbonate Precipitation (EICP) have gained attention in recent years for mainly three fields of application, (i) the targeted sealing of leakage pathways, (ii) the improvement of soils and other porous media, as well as (iii) their use in sustainable building materials, among other applications \citep{Smirnova2023,Zhang2023,Jimenez-Martinez2022,Alonso2019,Mujah2017,Umar2016,Phillips2013}. 
The engineered EICP process employs essentially two reactions.
The first is the irreversible hydrolysis of urea into ammonium and inorganic carbon, also called ureolysis. At low temperatures, as they normally prevail, the ureolysis needs to be catalyzed by the enzyme urease,
\begin{align}\label{eq:hydrolysisurea}
\ce{CO(NH2)2 + 3H2O ->[urease] 2NH4+ + HCO3- + OH-}.
\end{align}
As a result of the ureolysis, the pH increases and triggers the second reaction, the precipitation of calcium carbonate \citep{Jimenez-Martinez2022, Phillips2013},
\begin{align}\label{eq:precipitationcc}
\ce{HCO3- + Ca^{2+} -> CaCO3 v + H+}.
\end{align}
The ureolysis is assumed to be the slower reaction of the two and thus determines the overall precipitation rate \citep{van2009ground}.
A large influence is the inactivation of the available urease based on temperature or added chemicals \citep{feder2021temperature}.

In order to be able to control EICP processes, it is necessary to guarantee a sufficiently consistent stability, which is reflected by a homogeneous precipitation of calcium carbonate throughout the porous medium.
This is made more difficult by the multitude of factors that influence the amount and distribution of the precipitate, such as temperature, the inactivation of urease, and the dependence of the reaction rates on the concentrations of reactants, as well as the presence of other compounds in the solutions or even the grain and thus the pore size distribution of the porous medium \citep{feder2021temperature,machado2022crystallization, Konstantinou2023}.
The EICP process also has a much faster reaction rate than other biomineralization variants, such as Microbially Induced Calcium carbonate precipitation (MICP) \citep{saif2022advances}.
Since EICP is often carried out under forced flow-through conditions to ensure the supply of reactants, preferential flow paths can start forming, in particular due to initial heterogeneity of the porous medium, while the further evolution of pathways depends on a number of factors and boundary conditions.
We build models to gain a better understanding of the influence and interplay of these different factors.

In this work, we focus on the appearance of preferential flow paths during the mentioned forced flow-through operation of EICP. 
The formation of preferential flow paths has also been observed for abiotic calcium carbonate precipitation in fractures \citep{Jones2016}.
Flow paths occur first locally, triggered by heterogeneity in the pore-space distribution, which is typically not captured by the parameter porosity in REV-scale models \citep{Hommel2018Review}.
This can result in locally varying velocities and amounts of precipitate throughout the domain. This offers the flow to take the paths of least resistance, i.e., preferential flow paths.
Inside these flow paths, the velocity is higher than in their surroundings. 
On the one hand, this will cause more shear within a flow path, on the other hand, the flow path serves as a highway for supply with reactants as long as they were not depleted further upstream.
More shear means that precipitates along the path potentially get detached and transported away, while they can reattach at the edges of the path, where the velocity is lower.
Further, the ample reactant supply by the flow path promotes precipitation in its vicinity, while the fast velocity limits the residence time, and thus the time for precipitation in the flow path itself.
Both of these processes reduce the flow outside the flow paths even more, leading to a further increase in velocity in the paths, which makes the flow paths self-enhancing.
Even without shear and reattachment observed in \citet{weinhardt2022spatiotemporal}, calcium carbonate precipitation may still result in self-enhancing flow paths as in \citet{Jones2016}.
In addition, nucleation of calcite crystals may occur preferentially in low velocity areas, which will further enhance flow path formation \citep{Liu2023}. 
If the pressure becomes too high, the precipitates at the flow path edges can be flooded away, and the flow path may reform in a different manner.
From this behaviour we expect that two aspects are vital for the modeling of flow paths, (i) the initial heterogeneity in the pore space and (ii) a mechanism that allows the precipitates to detach and reattach.

Commonly, the expected product of EICP is assumed to be calcite, the most stable of the six calcium carbonate polymorphs.
In descending order of stability the polymorphs of calcium carbonate are calcite, aragonite, vaterite, monohydrocalcite, ikaite and amorphous calcium carbonate (ACC)\citep{de1998surface,el2013effects}.
Calcite is the polymorph most often observed in post-experiment sampling for MICP, which is closely related to EICP, \citep{Liu2023, Phillips2015, Mitchell2013}, although vaterite and amorphous calcium carbonate were also observed \citep{Liu2023, zambare2020mineralogy}.
For high supersaturation, meta-stable precursor minerals can appear and recrystallize as more stable variations \citep{kim2020assessing,shen2006properties,rodriguez2011kinetics}.
In EICP experiments, the polymorphs calcite, vaterite, and ACC have been observed, with calcite as the most common final result \citep{Xiao2024, weinhardt2022spatiotemporal}.
Using EICP by jack-bean (\textit{canavalia ensiformis}) urease to produce various calcium carbonate polymorphs, \citet{Sondi2001} found that when the solution was not agitated, ACC precipitated first, then recrystallized as vaterite and ultimately transformed to calcite. 
When the solution was stirred, \citet{Sondi2001} observed no ACC, but small vaterite precipitates transforming ultimately to calcite.
ACC is observed as a precursor phase that transforms into more stable polymorphs over time in MICP \citep{Liu2023}.
In general, the mechanisms of polymorph selection in urease-aided calcium carbonate precipitation are not yet fully understood \citep{Krajewska2017}.
Generally, ACC is very unstable and will transform into one of the more stable polymorphs within seconds. It has been observed, however, that in some crustaceans stabilized forms of ACC exist \citep{addadi2003taking}.
There is no known case of stable ACC and calcite coexisting in the same space.
Depending on the synthesis method and conditions, different forms of ACC were observed \citep{shen2006properties}.
The lifetime of ACC is highly dependent on the pH value and the polymorph, into which it recrystallizes, depends on temperature \citep{shen2006properties,rodriguez2011kinetics}.
Due to its high instability, ex-situ methods cannot be used to measure the lifetime of ACC \citep{shen2006properties}.
Properties of additives and its own solubility also influence the stability of ACC.
These effects make it almost impossible to obtain widely usable properties of ACC based on experimental results.

Modeling MICP and EICP assuming all calcium carbonate precipitates purely as calcite has shown to give good agreement with experimental results \citep{Feng2022, Wolff2021, Landa-Marban2021, hommel2020numerical, Hommel2015}. 
However, it also implies the assumption that calcite is immobile \citep{Feng2022, Landa-Marban2021, hommel2020numerical}.
In contrast, ACC has been observed in experiments to be mobile and appears as a precursor in EICP experiments \citep{Liu2023, Krajewska2017, Sondi2001}.
The formation of flow paths has also been observed in combination with ACC for EICP \citep{weinhardt2022spatiotemporal}. 
Thus, for the conceptual model of this study, we hypothesize ACC as a potentially mobile precipitate that can detach and reattach.
In addition, there is no unique possible description of the detachment and attachment processes for precipitates.

In this work, we make use of Bayesian approaches to consider the conceptual uncertainty of our hypothesis regarding the physical processes and to evaluate its plausibility.
In a Bayesian approach, unknown parameters are characterized as random variables with probability distributions that encode their uncertainty and take into account expert opinion. 
The influence of this uncertainty on the model can be analyzed by with sensitivity analysis and can be taken into account during inverse parameter estimation, i.e. Bayesian inference\citep{valdez2021foam,ranaee2022sensitivity,mohammadi2021surrogate}.
The uncertainty from having multiple options to formulate and build a model, i.e., conceptual uncertainty, can be considered by comparing the variations of models against each other in model validation or multi-model comparison.
Bayesian model justifiability analysis is a method for Bayesian multi-model comparison that compares a set of models against each other based on 'synthetic data', that is generated from the models.
This method is geared at identifying similarities between competing models, and it considers not only the approximation accuracy of each model, but also asks how much model complexity is justified by the available data \citep{schoniger2015finding}.
Recent extensions to include experimental data as a competing model allow to also gain understanding about the overall match between the set of models and data~\citep{scheurer2021surrogate}.

Bayesian model comparison approaches make use of brute-force calculation of likelihoods and posterior distributions, which require a large set of samples.
To make these evaluations feasible in terms of computational effort, the full models can be replaced with easy-to-evaluate surrogate models\citep{crevillen2019uncertainty,gadd2019surrogate,scheurer2021surrogate,mohammadi2021surrogate,ranaee2022sensitivity}.
Such surrogate models are built from a small set of model runs and specifically constructed to be cheap to evaluate.
Commonly used types include Neural Networks (NN) \citep{mcculloch1943logical}, Gaussian Process Emulators (GPE) \citep{williams2006gaussian} and Polynomial Chaos Expansions (PCE) \citep{wiener1938homogeneous}.
These can be used individually or in combination, depending on the amount of available data and the properties of the model they are fitted to \citep{bradford2021combining,kohlhaas2023gaussian,oladyshkin2023deep,yao2023deep}.
We use here the arbitrary Polynomial Chaos Expansion (aPCE), which expands the scope of the PCE to sample-based prior definitions \citep{oladyshkin2012data}.
Sparse training methods help reduce the number of polynomial terms with less influence\citep{luethen2021sparse,tipping2001sparse,tipping2003fast,meng2019efficient,burkner2023fully}. 

Section \ref{sec:model} briefly reviews the existing EICP model, which we then extend to account for ACC.
We introduce two variations of the extended model that differ in the assumptions about the state that ACC precipitates in.
In section \ref{sec:Bayes}, we present the chosen aPCE surrogate model, and the Bayesian approaches that we apply in this study for model validation and comparison.
Section \ref{sec:setup} contains a description of a microfluidic experiment from \citet{weinhardt2022spatiotemporal} that has resulted in preferential flow, the setup of the model with initial and boundary conditions and the setup of the surrogate model including prior distributions for the unknown model parameters.
The model results are presented and discussed in section \ref{sec:results} with respect to the general model behaviour, the appearance of flow paths for different setups and the influence of the uncertain parameters. 
The model variations introduced in section \ref{sec:model} are compared against each other with Bayesian model analysis approaches.

\section{Modeling enzymatically induced calcium carbonate precipitation}\label{sec:model}
\begin{figure} [tb]\centering
 \subfigure[]{\includegraphics[width=0.35\textwidth]{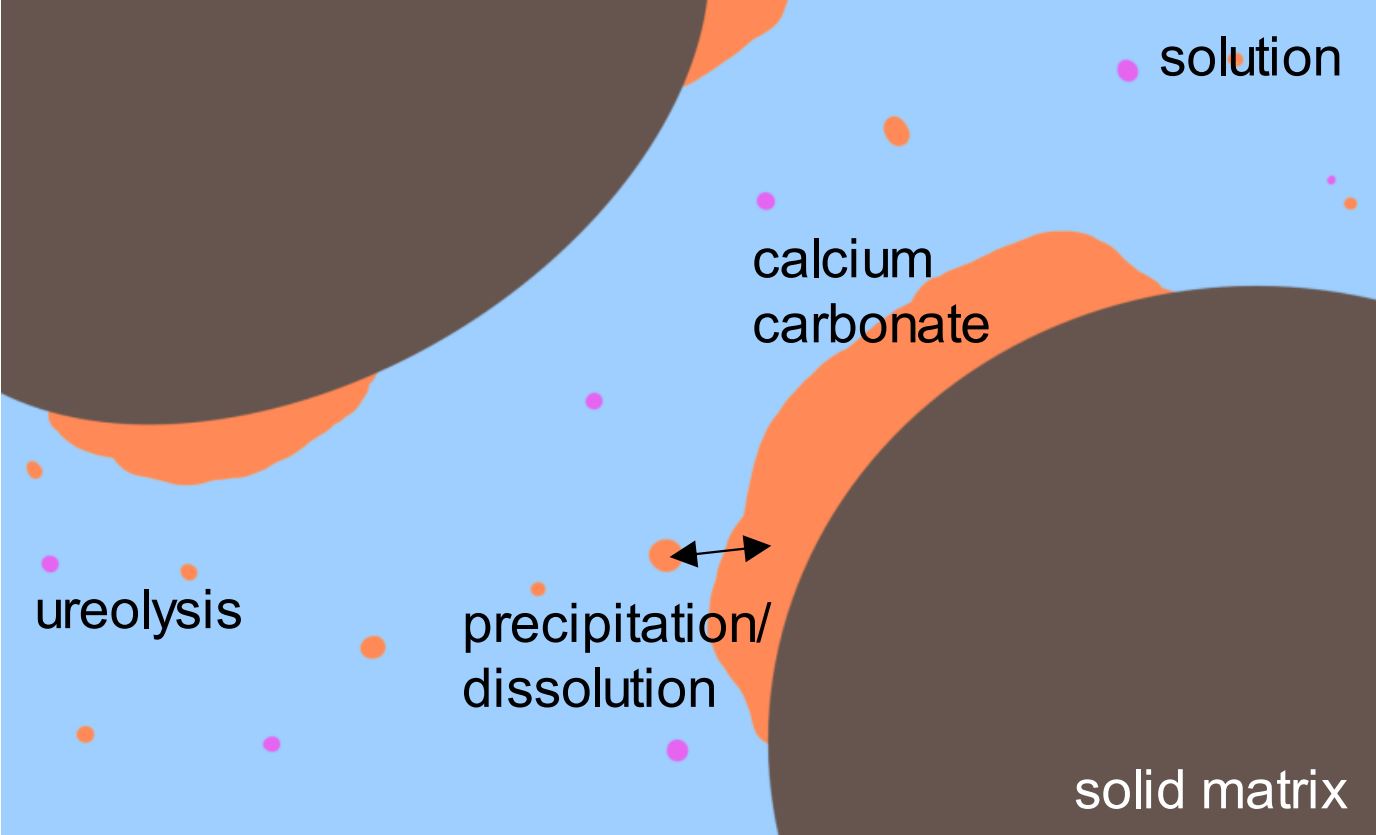}}
 \subfigure[]{\includegraphics[width=0.35\textwidth]{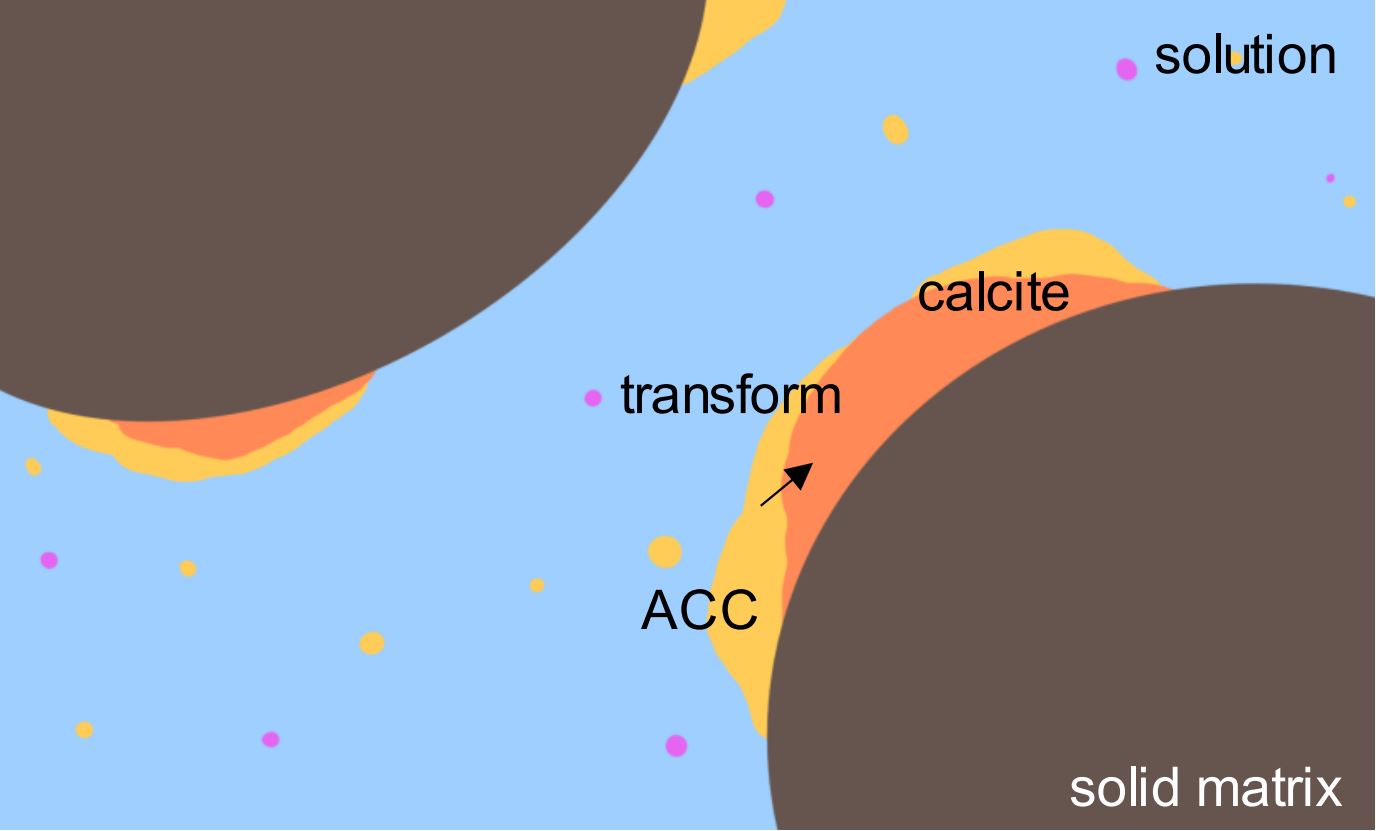}}
\caption{Schematic pore-scale visualization of EICP accounting for (a) only calcium carbonate as a whole and (b) ACC and calcite as distinct polymorphs of calcium carbonate and the transformation of ACC into calcite.} \label{fig:EICP}
\end{figure}
The model used in this study is based on the EICP model presented in \citet{hommel2020numerical} with further adaptions to the specific conceptual assumptions of this study.

A simplified illustration of the presumed processes on the pore scale is shown in figure \ref{fig:EICP}, while the model itself is formulated on the averaged scale of an REV. Fig.~\ref{fig:EICP} (left) displays only a single precipitate type; in contrast, on the right figure, calcite and ACC exist separately with ACC transforming over time into calcite.
The following assumptions summarize the extension of the EICP model by \citet{hommel2020numerical}. 
\begin{itemize}
    \item ACC may be present in addition to calcite.
    \item All calcium carbonate precipitates as ACC and transforms into calcite over time.
    \item ACC can exist both in an attached or a detached state, and can switch between the two by attachment and detachment processes.
    \item There is no dissolution of precipitated calcium carbonate.
\end{itemize}

\subsection{Basis EICP model}\label{sec:EICPmodel}
We summarize briefly the EICP model presented in \citet{hommel2020numerical} with a particular focus on the sink and source terms that will be extended in section \ref{sec:extendModel}.

The EICP model considers two-phase, multicomponent, non-isothermal reactive Darcy flow and transport.
The model distinguishes mobile components $\kappa$ from immobile components $\lambda$.
The mobile components are water $\ce{w}$, dissolved inorganic carbon $\ce{C_{tot}}$, sodium $\ce{Na}$, chloride $\ce{Cl}$, calcium $\ce{Ca}$, urea $\ce{u}$, ammonium and ammonia $\ce{N_{tot}}$, and suspended urease $\ce{su}$.
Immobile components are calcite $\ce{c}$ and adsorbed urease $\ce{au}$.
The component balance equations are solved for the aqueous phase pressure $p_\mathrm{w}$ and for the mole fractions $x^\kappa_\mathrm{w}$ of component $\kappa$ in the water phase as the primary variables.
For the solid phase, the volume fractions $\phi_\lambda$ are used instead of mole fractions.
Whenever both fluid phases are present in the same control volume, the gas phase saturation $S_\mathrm{g}=\frac{\mathrm{gas volume}}{\mathrm{total pore volume}}$ replaces the mole fraction $x_\mathrm{w}^{\mathrm{C}_\mathrm{tot}}$ of total inorganic carbon in water.
The balance equations are formulated as follows:
\begin{align}\label{eq:balanceEICP}
    \sum_\alpha(\frac{\partial}{\partial t}(\phi\rho_\alpha x_\alpha^\kappa S_\alpha)+\nabla\cdot(\rho_\alpha x_\alpha^\kappa v_\alpha)-\nabla\cdot(\rho_\alpha D^\kappa_\mathrm{pm}\nabla x_\alpha^\kappa))=q^\kappa,\\
    \frac{\partial}{\partial t}(\phi_\lambda\rho_\lambda)=q^\lambda,
\end{align}
where $t$ is time, $\phi\rho_\alpha$ the molar density of phase $\alpha$, $x_\alpha^\kappa$ the mole fraction of component $\kappa$ in phase $\alpha$, $v_\alpha$ the Darcy velocity of phase $\alpha$, $D^\kappa_\mathrm{pm}$ the diffusion coefficient of component $\kappa$ in phase $\alpha$, and $q^\kappa$ the source or sink term for component $\kappa$.

All reactive and mass transfer processes are incorporated as component-specific source and sink terms.
The reactive processes include two temperature-dependent reactions, urease inactivation and enzymatically catalyzed ureolysis, and pH dependent dissociation reactions.
Mass transfer between the phases is obtained via mutual dissolution of water and $\ce{CO2}$ in the gaseous or aqueous phases, adsorption and desorption of urease, and dissolution of calcium carbonate.

The source and sink terms are given by 
\begin{align*} 
    q^\mathrm{w} &= 0,  &  q^\mathrm{C}_{tot} &= r_\mathrm{diss}-r_\mathrm{prec}+r_\mathrm{u},\\
    q^\mathrm{Na} &= 0,  &  q^\mathrm{Ca} &= r_\mathrm{diss}-r_\mathrm{prec},\\
    q^\mathrm{Cl} &= 0,  & q^\mathrm{su} &= r_\mathrm{d}-r_\mathrm{a}-r_\mathrm{i}^\mathrm{su}, \\
    q^\mathrm{N}_\mathrm{tot} &= 2r_\mathrm{u}, &  q^\mathrm{au} &= -r_\mathrm{d}+r_\mathrm{a}-r_i^\mathrm{au},\\
    q^\mathrm{u} &= -r_\mathrm{u},  &  q^\mathrm{c} &= r_\mathrm{prec}-r_\mathrm{diss},
\end{align*}
with $r_\mathrm{u}$ the ureolysis rate, $r_\mathrm{diss}$ and $r_\mathrm{prec}$ the calcite dissolution and precipitation rates,  $r_\mathrm{d}$ and $r_\mathrm{a}$ the urease desorption and adsorption rates and $r_\mathrm{i}^\mathrm{su}$ and $r_\mathrm{i}^\mathrm{au}$ the inactivation rates for suspended and adsorbed urease.

The calcium carbonate precipitation rate $r_{prec}$ is given by 
\begin{align}\label{eq:rprec}
    r_\mathrm{prec} &= k_\mathrm{prec}A_\mathrm{sw}(\Omega-1)^{n_\mathrm{prec}},\\
    A_\mathrm{sw} &= A_\mathrm{sw,0}(1-\frac{\phi_\mathrm{c}}{\phi_0})^{\frac{2}{3}}. \label{eq:AswOld}
\end{align}
We note that $r_\mathrm{prec}$ is calculated only if $\Omega$, the saturation state with respect to calcium carbonate, is greater than one.
Otherwise $r_\mathrm{prec}$ is set to zero.
The precipitation rate coefficients $k_\mathrm{prec}$ and $n_\mathrm{prec}$ as well as the surface area $A_\mathrm{sw}$ are given in dependence of the initial surface area $A_\mathrm{sw,0}$, the precipitate volume fraction $\phi_\mathrm{c}$, and the initial porosity $\phi_0$.
For the specific reaction rates, see \citet{hommel2020numerical}.

\subsection{Extended EICP model}\label{sec:extendModel}
We extend the model by two additional components, i.e., attached and detached ACC, with corresponding source and sink terms $q^\mathrm{ACC}_\mathrm{a}$ and $q^\mathrm{ACC}_\mathrm{d}$.
We thus introduce ACC as a metastable calcium carbonate polymorph that may be mobile, while calcite remains immobile.
The source and sink terms of the original model, as discussed in section \ref{sec:EICPmodel}, are now as follows:
\begin{align*}
    q^\mathrm{C}_\mathrm{tot} &= -r_\mathrm{prec}+r_\mathrm{u},\\
    q^\mathrm{Ca} &= -r_\mathrm{prec},\\
    q^\mathrm{c} &= r_\mathrm{prec}^\mathrm{c}+r_\mathrm{ACC}2c.\\
\end{align*}
We recall the assumption that we do not consider dissolution of calcium carbonates in this work, contrary to the original model by \citet{hommel2020numerical}.

Two possible states for newly precipitated ACC are considered, i.e., fully attached or fully detached.
We create two separate models based on this distinction, which differ according to their source term for ACC precipitation.
Precipitation of ACC was observed both as attached (heterogeneous nucleation) and as detached (homogeneous nucleation) under flow conditions in microfluidic investigations \citep{Xiao2024}.
For the attached precipitation model variant ($M_\mathrm{attached}$), the precipitated amount of ACC is added to $q^\mathrm{ACC}_\mathrm{a}$,
\begin{align*}
    q^\mathrm{ACC}_\mathrm{a} &= r_\mathrm{prec}^\mathrm{ACC}-r_\mathrm{d}\mathrm{ACC}+r_\mathrm{a}^\mathrm{ACC} -r_\mathrm{ACC2c},\\
    q^\mathrm{ACC}_\mathrm{d} &= r_\mathrm{d}^\mathrm{ACC}-r_\mathrm{a}^\mathrm{ACC}.
\end{align*}
For the detached precipitation model variant ($M_{detached}$), it is added to $q^\mathrm{ACC}_\mathrm{d}$,
\begin{align*}
    q^\mathrm{ACC}_\mathrm{a} &= -r_\mathrm{d}^\mathrm{ACC}+r_\mathrm{a}^\mathrm{ACC} -r_\mathrm{ACC2c},\\
    q^{\mathrm{ACC}}_\mathrm{d} &= r_\mathrm{prec}^\mathrm{ACC} + r_\mathrm{d}^\mathrm{ACC}-r_\mathrm{a}^\mathrm{ACC}.
\end{align*}

The total calcium carbonate precipitation $r_\mathrm{prec}$ is calculated in the same manner as described in section \ref{sec:EICPmodel}, but all precipitation is allocated to ACC,
\begin{align}
    r_\mathrm{prec}^\mathrm{c} &= c_\mathrm{prec}\cdot r_\mathrm{prec}\frac{\phi^\mathrm{c}}{\phi^\mathrm{c}+\phi^\mathrm{ACC}}.
\end{align}

We update the specific interfacial area calculation from Eq.~\eqref{eq:AswOld}, as we have the geometric information available from the experiments of \citet{weinhardt2022spatiotemporal}, which inspired our setup. 
We assume a parabolic form for $A_\mathrm{sw}(\Bar{\phi})$ and fit it to the experimentally observed specific surface area for a given relative change in porosity $\Bar{\phi}$. 
Section \ref{ssec:surface} in the appendix details the calculation of the specific surface area based on the experiment of \citet{weinhardt2022spatiotemporal} as well as the parameter estimation procedure for the parabolic relationship.

We model the transformation from attached ACC into calcite as an exponential decay with half life $T_\frac{1}{2}$:
\begin{align}\label{eq:ACCtoCalcite}
    r_\mathrm{ACC2c} = \frac{\ln(2)}{T_\frac{1}{2}}\cdot  \phi^\mathrm{ACC}\cdot  \frac{\rho_\mathrm{ACC}}{M_\mathrm{ACC}}.
\end{align}

The attachment rate of ACC is calculated depending on the available specific surface area $A_\mathrm{sw}$.
The surface area estimation was performed as described in Section \ref{ssec:surface} in the appendix.
It is scaled by a coefficient $c_\mathrm{a}$,
\begin{align} \label{eq:ACCattach}
    r^\mathrm{ACC}_\mathrm{a} &= c_\mathrm{a} \cdot  A_\mathrm{sw} \cdot C^\mathrm{ACC}_w \cdot \frac{\phi S_\mathrm{w}}{M_\mathrm{ACC}}.
\end{align}

Here, $C^\mathrm{ACC}_w$ is the ACC mass concentration in the water phase, $S_\mathrm{w}$ the water phase saturation, and $M^\mathrm{ACC}$ the molar mass of ACC.
We chose the above rate equation for its simplicity. 
$A_\mathrm{sw}$ is included as a multiplier to $c_\mathrm{a}$ since the attachment of suspended ACC particles to the solid fundamentally occurs at the solid-water interfacial area.
Further, including $A_\mathrm{sw}$ allows to represent the heterogeneity of the sample-setup porous media.

The detachment rate of ACC depends on the amount of ACC in an REV, which can be calculated as $\phi_\mathrm{ACC} \frac{\rho_\mathrm{ACC}}{M_\mathrm{ACC}}$.
It is scaled by a coefficient $c_\mathrm{d}$, and we assume dependence on the square of the effective velocity, $v_\mathrm{eff}^2$, 
\begin{align} \label{eq:ACCdetach}
    r^\mathrm{ACC}_\mathrm{d} &= c_\mathrm{d} \cdot  v_\mathrm{eff}^2 \cdot  \phi_\mathrm{ACC}\cdot  \frac{\rho_\mathrm{ACC}}{M_\mathrm{ACC}}.
\end{align}
We hypothesize that detachment of ACC occurs due to drag forces, which are proportional to the square of the velocity of the fluid having to flow around an obstacle. 
On the REV scale however, the exact velocity profiles within pores are unknown sub-scale information. 
We use $v_\mathrm{eff}$ instead of the Darcy velocity $v$ to account for the lack of detailed velocity information and to ensure a stable regime at which no detachment occurs. 
We define the effective velocity so that it relates the Darcy velocity $v$ to a minimum velocity $v_\mathrm{Crit}$ that should be surpassed for detachment to occur,
\begin{align}
    v_\mathrm{eff}^2=\mathrm{min}\left(0,(v-\mathrm{v}_\mathrm{Crit})\right)^2.
\end{align}

These equations give a set of four ACC-specific parameters for both model variations $M_\mathrm{attached}$ and $M_\mathrm{detached}$:
\begin{itemize}
    \item \text{$c_\mathrm{d}$}: Detachment constant
    \item \text{$c_\mathrm{a}$}: Attachment constant
    \item \text{$T_\frac{1}{2}$}: Half life of ACC 
    \item $v_\mathrm{Crit}$: Critical velocity for detachment

\end{itemize}

Introducing attached ACC as an additional solid phase, its volume fraction $\phi_\mathrm{ACC}$ has to be considered when calculating the current porosity $\phi$ in addition to the volume fraction of calcite $\phi_\mathrm{c}$ and the initial porosity $\phi_0$,
\begin{align}\label{eq:porosity}
    \phi = \phi_0 - \phi_\mathrm{ACC} - \phi_\mathrm{c}.
\end{align}

A further extension to the original model is the use of an anisotropic power law for the porosity-permeability relation,
\begin{align}\label{eq:poro-perm}
    K_x = \left(\frac{\phi}{\phi_0}\right)^{\eta_x} K_0, &  & 
    K_y = \left(\frac{\phi}{\phi_0}\right)^{\eta_y} K_0.
\end{align}
Here, $K_x$ and $K_y$ are the current permeabilities in the direction of the main flow and perpendicular to it, respectively, $K_0$ is the initial permeability, and  $\eta_x$ and $\eta_y$ are the exponents of the anisotropic power law in the direction of the main flow and perpendicular to it, respectively.
The exponent along the main flow direction, $\eta_x=2.516$ is calculated to match the initial and final porosities as well as permeabilities measured by \citet{weinhardt2022spatiotemporal} using Equation~\eqref{eq:poro-perm}, see also Section~\ref{sec:experiment}.
For the direction perpendicular to the main flow, there is no information available. 
However, the permeability perpendicular to the main flow direction is expected to be reduced much more due to the flow-path structure observed in the experiment. 
Thus, we set the exponent in this direction to $\eta_y=5\approx 2\times \eta_x$ to facilitate flow path formation.
Note that since our aim is to enable investigation of flow-path formation due to the induced precipitation, we can not explicitly represent the flow path as lower-dimensional objects as is commonly done for other structured porous media, e.g. fractures in \citet{Banshoya2023}, as this would require a-priory knowledge on where flow paths will eventually develop.
The higher the exponent in a power-law relation is chosen, the stronger the permeability changes for a given change in porosity \citep{Hommel2018Review}.

\section{Bayesian model analysis}\label{sec:Bayes}
To make model evaluations described in chapter \ref{sec:model} feasible despite their complexity, we apply Bayesian surrogate modeling, which enables efficient approximation of the original model’s output while significantly reducing the computational cost. 
This approach approximates model outputs by training surrogate models on a limited set of model evaluations. 

This chapter presents the type of surrogate model and the Bayesian methods that are used in this work.
Three aspects are differentiated: the measured data $D$ which has been generated by an underlying process for unknown 'true' parameters $\theta_{true}$, a model $M$ with $p$ uncertain inputs $\theta=(\theta_1,\dots,\theta_p)$ and model response $Y=M(\theta)$, and the surrogate $\bar{M}$ which takes the same inputs $\theta$ as the model. 
The surrogate is fitted to a set of input samples and model response $D_{exp}=\{\theta_t,Y_t\}$, called the training set.

\subsection{Bayesian sparse arbitrary polynomial chaos expansion}\label{sec:aPC}
In this work we use the arbitrary Polynomial Chaos Expansion (aPCE) \citep{oladyshkin2012data,oladyshkin2018incomplete}, a variation of the Polynomial Chaos Expansion (PCE) \citep{wiener1938homogeneous} that allows for data-driven input parameter distributions.
Indeed, let $(\Omega, \Sigma, \Gamma)$ be a probability space, with space of events $\Omega$, $\sigma$-Algebra $\Sigma$ and probability measure $\Gamma$.
Let $\theta=(\theta_1,\dots,\theta_p)\in\Omega^p$, $p\in\mathbb{N}$ be a random variable that describes an uncertainty and $Y$ a system response to be expressed in terms of $\theta$.
PCE expresses the system response $Y$ as an infinite sum of polynomials $\Psi_\alpha$ with expansion coefficients $c_\alpha$, 
\begin{align}
    Y \approxeq \bar{M}(\theta)=\sum_{\alpha\in \mathbb{N}^d}c_\alpha\Psi_\alpha(\theta).
\end{align}
For computational reasons, the sum is truncated to a set of multi-indexes $A \subset \mathbb{N}^d$. 
Different truncation schemes are available, here we use the standard total-degree truncation scheme,
\begin{align}
    \mathcal{A}^{d,p} = \{\alpha\in\mathbf{N}^{d}:|\alpha|\leq d\}.
\end{align}
A discussion on using hyperbolic truncation via the $q$-norm introduced in \citet{blatman2011adaptive} can be found in appendix \ref{app:sensitivity}.

The polynomials $\Psi_\alpha(\theta)$ are defined as the tensor product of univariate polynomials,
\begin{align}
    \Psi_\alpha(\theta)=\prod_{i=1}^d\psi_{\alpha_i}^{(i)}(\theta_i),
\end{align}
which should form an orthonormal basis for  Kronecker delta $\delta_{j,k}$ and $f_{\theta_i}(\theta_i)$ the $i$th input marginal distribution,
\begin{align}
    <\psi_j^{(i)},\psi_k^{(i)}>=\int_\Omega\psi_j(\theta_i)\psi_k(\theta_i)f_{\theta_i}(\theta_i)d\theta_i=\delta_{j,k}.
\end{align}

The univariate polynomials can be chosen following different routes.
Arbitrary Polynomial Chaos Expansion (aPCE) calculates the orthogonal polynomials for any arbitrary random variable $\theta$ based on data-driven distributions \citep{oladyshkin2012data,oladyshkin2018incomplete} by defining a moment-based method of constructing the polynomials from the Hankel matrix of moments. 
Using aPCE on a random variable $\theta$ that is included in the Askey scheme leads to the same set of polynomials as are described in the scheme. 
A more complete explanation of the algorithm can be found in \citet{oladyshkin2018incomplete}.

We further reduce the number of needed training samples using a Bayesian sparse training method. Indeed, sparse learning methods help to further improve the surrogate approximation for a small number of training points by ensuring that a higher number of coefficients $c_\alpha$ become $0$.
In this work we make use of the Bayesian sparse learning method FastARD \citep{tipping2001sparse,tipping2003fast,mohammadi2023surrogate}.
In addition to sparsity, FastARD also provides Gaussian properties to the aPCE.
This means that, similarly to Gaussian Process Emulators, each evaluation of an aPCE trained with FastARD returns not only an approximation of the model output, but also the uncertainty associated with each surrogate prediction.
These values are used in section \ref{sec:MJA} to compensate for comparing surrogates instead of the underlying models.

FastARD imposes sparsity inducing priors on the coefficients of the aPCE. 
It defines priors on the coefficients $c$ with hyperparameters $\alpha$ that describe the precision the prior has over $c$.
For $\bar{A}=\mathrm{card} (A)$, define
\begin{align}
    P(c|\alpha)=\prod_{m=1}^{\bar{A}} \mathcal{N}(c_m|0,\alpha_m^{-1}).
\end{align}
The form of the prior is responsible for the sparsity properties \citep{tipping2001sparse}.
Setting $\alpha_i=\infty$ means that the corresponding basis function $\Psi_i$ is not used in the surrogate approximation.
Posteriors for $c$ are found via type-II maximum likelihood \citep{berger2013statistical}.
The likelihood is given as 
\begin{align}
    P(Y|X,\alpha,\beta)=\int P(Y|X,c,\beta)P(c|\alpha)dc,
\end{align}
under the assumption that the approximation error of the PCE is a zero-mean Gaussian $\epsilon\sim \mathcal{N}(0,\beta^{-1})$ with precision $\beta$.
Based on \citet{faul2001analysis} a unique maximum for the log-likelihood exists.
Starting with an empty expansion, basis functions are iteratively included to increase the marginal likelihood while modifying the weights $c_\alpha$.
For an in-depth description of the calculation, see \citet{tipping2003fast}.

\subsection{Sobol' indices} \label{ssec:sobol}
Sensitivity analysis of the surrogate with respect to the parameters $\theta$ gives an understanding of the importance of each input parameter.
Sobol' indices are a widely used approach for sensitivity analysis, particularly suited for capturing parameter influence in models with non-linear and non-monotonic response surfaces \citep{sobol1993sensitivity}.

As the parameters in the random vector $\theta=(\theta_1,...,\theta_p)$ are assumed to be mutually independent, the variance $\sigma^2(Y)$ of the model response can be decomposed into contributions from each uncertain parameter $\theta_i$ \citep{efron1981jackknife}.
For aPCE, the orthonormality of the polynomial basis allows to calculate the Sobol' indices directly from the coefficients $c_\alpha$ \citep{sudret2008global}.
The Sobol' index $S^i$ contains the contribution of a set of parameters $i$ to the response variance without considering cross-contributions from other parameters,
\begin{align}
    S^i=\frac{\sum_{\alpha\in A_i}c_\alpha^2}{\sum_{\alpha\in A}c_\alpha^2}\text{ ,for }A_i^T={\alpha\in A: \alpha_j>0,a_{j\neq i}=0},
\end{align}
for $A_i$ all possible polynomial degrees that contain only the values $i$.

Additionally, the Total Sobol' Index calculates the total contribution of a single parameter $i$ including all cross-combinations with other parameters,
\begin{align}
    S_{T}^i=\frac{\sum_{\alpha\in A_i^T}c_\alpha^2}{\sum_{\alpha\in A}c_\alpha^2}\text{ , for }A_i^T={\alpha\in A: \alpha_j>0}.
\end{align}
Note here that while the complete set of Sobol' indices will add to $1$, the Total Sobol' indices can be larger than 1 if the contribution from combination terms is not negligible.

\subsection{Model justifiability analysis}\label{sec:MJA} 
\begin{figure}[tb]\centering
    \subfigure[]{\includegraphics[width=0.5\linewidth]{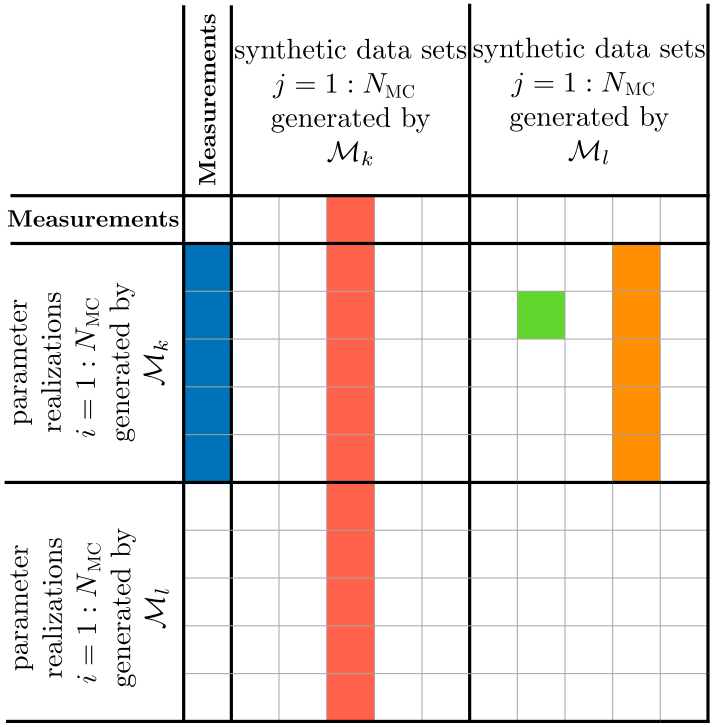}}
    \caption{Schematic visualization of the construction of the confusion matrix with extension by data $D$, image from \cite{mohammadi2023surrogate}.}\label{fig:confusionmatrix}
\end{figure}
To compare the model variants introduced in section \ref{sec:model}, we apply Bayesian multi-model comparison via model justifiability analysis, which ranks a set of competing models against each other with the help of a model confusion matrix.
This method helps determine the model that best represents the physical processes underlying the observed results and quantifies the similarity between models.
The confusion matrix calculates the posterior probability weights $P(M_k|D)$ for each model $M_k$, indicating the likelihood that model is the most accurate representation of the synthetic data $D$.
Synthetic data $D$ is generated from each model variant and compared across the full model set, with posterior weights $P(M_k|D)$ determined via Bayesian Model Selection \citep{hinne2020conceptual}.

The posterior model weights $P(M_k|D)$ are derived using Bayes' theorem, which allows us to interpret each weight as the probability that model $M_k$ represents the data the best among all models considered:

\begin{align}
    P(M_k|D)=\frac{P(D|M_k)P(M_k)}{\sum_{i=1}^{N_m}P(D|M_i)P(M_i)},
\end{align}
where $P(M_k)$ is the prior model weight and $P(D|M_k)$ is the Bayesian Model Evidence (BME). In the current work, we set the values of prior $P(M_k)$ to a neutral choice $P(M_k)=\frac{1}{N_m}$, assuming all models as equally likely.
The BME value $P(D|M_k)$ can be computed as  
\begin{align}
    P(D|M_k)=\int_{\Omega_k}P(D|M_k,\theta)P(\theta|M_k)d\theta,
\end{align}
where we choose a Gaussian likelihood function with zero mean to quantify how well model evaluations $M_k(\theta)$ match the measured data $D$,
\begin{align}
    P(D|M_k,\theta)=(2\pi)^{-N_s/2}|R|^{-1/2}\exp(-\frac{1}{2}(D-M_k(\theta))^TR^{-1}(D-M_k(\theta))).
\end{align}
Here, $R$ is the covariance matrix of the measurement error $\epsilon$ of the measured data $D$.

The construction of the confusion matrix is visualized in figure \ref{fig:confusionmatrix}.
Each cell, colored in green, represents the likelihood $P(M_{l,i}|M_{k,j})$ of a specific model realization $M_{l,i}$ given a specific value $M_{k,j}$ of synthetic data.
The BME of model $M_l$ for a synthetic data value is calculated from the individually calculated cells (in orange), and can be normed by the sum of BME values for the specific data value (in red) to get posterior weights for each data value. Considering the surrogate representation of each model, integration over the statistical space can be estimated via a Monte Carlo approach. Consequently, the final posterior weights $W_{l|k}^\mathrm{post}$ are given by each full box of likelihoods for model $M_k$ based on model $M_l$,
\begin{align}
    W_{l|k}^\mathrm{post}=\frac{1}{N_\mathrm{MC}}\sum^{N_\mathrm{MC}}_{j=1}P(M_l|M_{k,j})=\frac{1}{N_\mathrm{MC}}\sum^{N_\mathrm{MC}}_{j=1}\sum^{N_\mathrm{MC}}_{i=1}p(M_{l,i}|M_{k,j}).
\end{align}

The model confusion matrix can be extended by adding the measured data $D$ as a competing model, which increases the matrix size to $(N_m+1)\times(N_m+1)$ \citep{scheurer2021surrogate}.
The approximation errors between surrogate and model can be taken into account by introducing surrogate-specific corrective weights \citep{mohammadi2018bayesian,scheurer2021surrogate}.
The corrected posterior model weights are given by
\begin{align}
    W^\mathrm{post}&=\frac{1}{N_\mathrm{MC}}\sum_{j=1}^{N_\mathrm{MC}}P(M_k|M_l)=\frac{1}{N_\mathrm{MC}}\sum_{j=1}^{N_\mathrm{MC}}P(\bar{M}_k|\bar{M}_l)W_{\bar{M}_k}W_{\bar{M}_l},
\end{align}
where the weight for surrogate $\bar{M}_k$ is defined as
\begin{align}
    W_{\bar{M}_k}&=\int_{\theta_l}P(M_k|\bar{M}_k,\theta)P(\theta|\bar{M}_l\bar{M}_k)d\theta.
\end{align}
We follow \citet{scheurer2021surrogate} and choose a Gaussian likelihood for the comparison of model and surrogate model outputs with a covariance $S$ that describes the surrogate approximation errors,
\begin{align}
    P(M_k|\bar{M}_k,\theta)=(2\pi)^{-N_s/2}|S|^{-1/2}\exp(-\frac{1}{2}(M_k-\bar{M}_k(\theta))^TS^{-1}(M_k-\bar{M}_k(\theta))).
\end{align}
The corrective weights accounting for surrogate error \citep{scheurer2021surrogate} are ideally estimated with Monte Carlo integration from additional model runs, but if this is not feasible the $N_\mathrm{train}$ evaluations in the training set can also be used,
\begin{align}
    W_{\bar{M}_k}\approx \sum_{i=1}^{N_\mathrm{train}}P(M_k|\bar{M}_k,\theta_i^\cdot )P(\theta_i^\cdot |\bar{M}_l\bar{M}_k)d\theta).
\end{align}

\section{Setup of model and evaluations}\label{sec:setup}
This study was inspired by a microfluidic EICP experiment observing a preferential flow path. 
The experiment is briefly summarized as a basis to describe the setups for both the numerical model for EICP, see section~\ref{sec:EICPmodel}, and the surrogate models used for model Bayesian analysis and multi-model comparison. 

\subsection{Experiment 'Quasi-2D-1'}\label{sec:experiment}
The experimental basis of this study are the results of experiment 'Quasi-2D-1' of \citet{weinhardt2022spatiotemporal}, a microfluidic EICP experiment, which was restarted after leakage and clogging issues in the inlet of the microfluidic cell.

\begin{figure} [htb]
	\centering
	\includegraphics[width=0.65\linewidth]{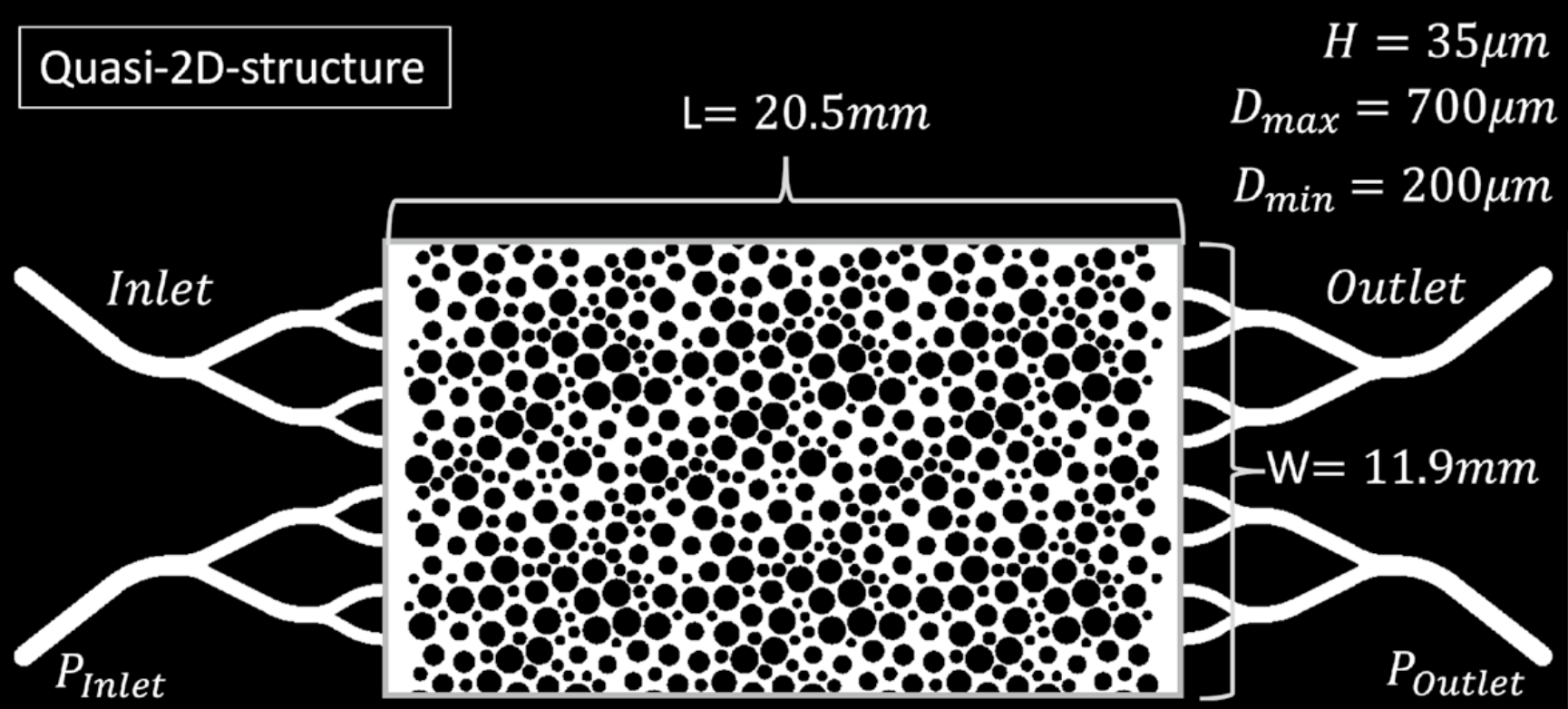}
	\caption{Setup of the microfluidic cell, image from \citet{weinhardt2022spatiotemporal} (Licence: CC BY-NC-ND).} \label{fig:cell}
\end{figure}

Experiment 'Quasi-2D-1' was performed in a borosilicate glass microfluidic cell fabricated by Micronit\textcopyright{} with a porous domain of $20.5\times11.9\times0.035\mathrm{ mm}^3$.
The solid matrix in the porous domain is given by pillars of various sizes with diameters ranging from $200$ to $700\text{ $\mu$m}$.
The resulting initial pore space is fully connected and has an extension in the through-plane of $35\text{ $\mu$m}$.
The porous domain of the flow cell is visualized in figure \ref{fig:cell} with the solid matrix in black and the initial pore space in white.

The cell is connected to two channels on the left and rights sides that both connect to distribution channels.
The outlet tube is connected to a reservoir with a constant head elevated $10\mathrm{ cm}$ above the cell.
Pressure sensors were connected to one channel on each side, in parallel to those channels used as the inlet and outlet to improve the reliability of the pressure difference measurements.

Two reactive solutions were prepared in accordance with \citet{weinhardt2021experimental}, Solution~$1$ containing urea and calcium chloride dihydrate at equimolar concentrations of $c=\frac{1}{3}\frac{\mathrm{mol}}{\mathrm{l}}$, and Solution~$2$ containing urease extracted from Jack-Bean meal.
The specific preparation process of the second solution is described in \citet{weinhardt2022spatiotemporal}.

During the experiment, the pressure differences and corresponding flow were logged.
Images of the cell were taken using optical microscopy with a resolution of $2280\times 1320$ pixel (px), resulting in an approximate resolution of $l_{px}=0.009\frac{\mathrm{mm}}{\mathrm{px}}$.
The precipitated volumes were estimated assuming frustum shapes of the precipitates, which has been validated based on $\mu$-XRCT scans described in \citep{weinhardt2022spatiotemporal}. 
Based on this reconstruction of the calcium carbonate precipitates, the pixels were divided into empty pixels with volume $V_{px}=0$, filled pixels with volume $V_{px}=1$, and border pixels with volume $0<V_{px}<1$.
The corresponding porosities were obtained as $\phi = \frac{V_\mathrm{void}}{V_\mathrm{total}}$

The experiment was run for a total time of $T_\mathrm{end}=4\mathrm{h}=14400$~s with constant flow rate $Q = 0.04\frac{\mu \mathrm{l}}{\mathrm{s}}$.
As it is a restarted experiment, the cell already contained precipitated calcite at $t=0$~s.

\begin{figure} [tb]\centering
 \subfigure[$t=0$h]{\includegraphics[width=0.33\textwidth]{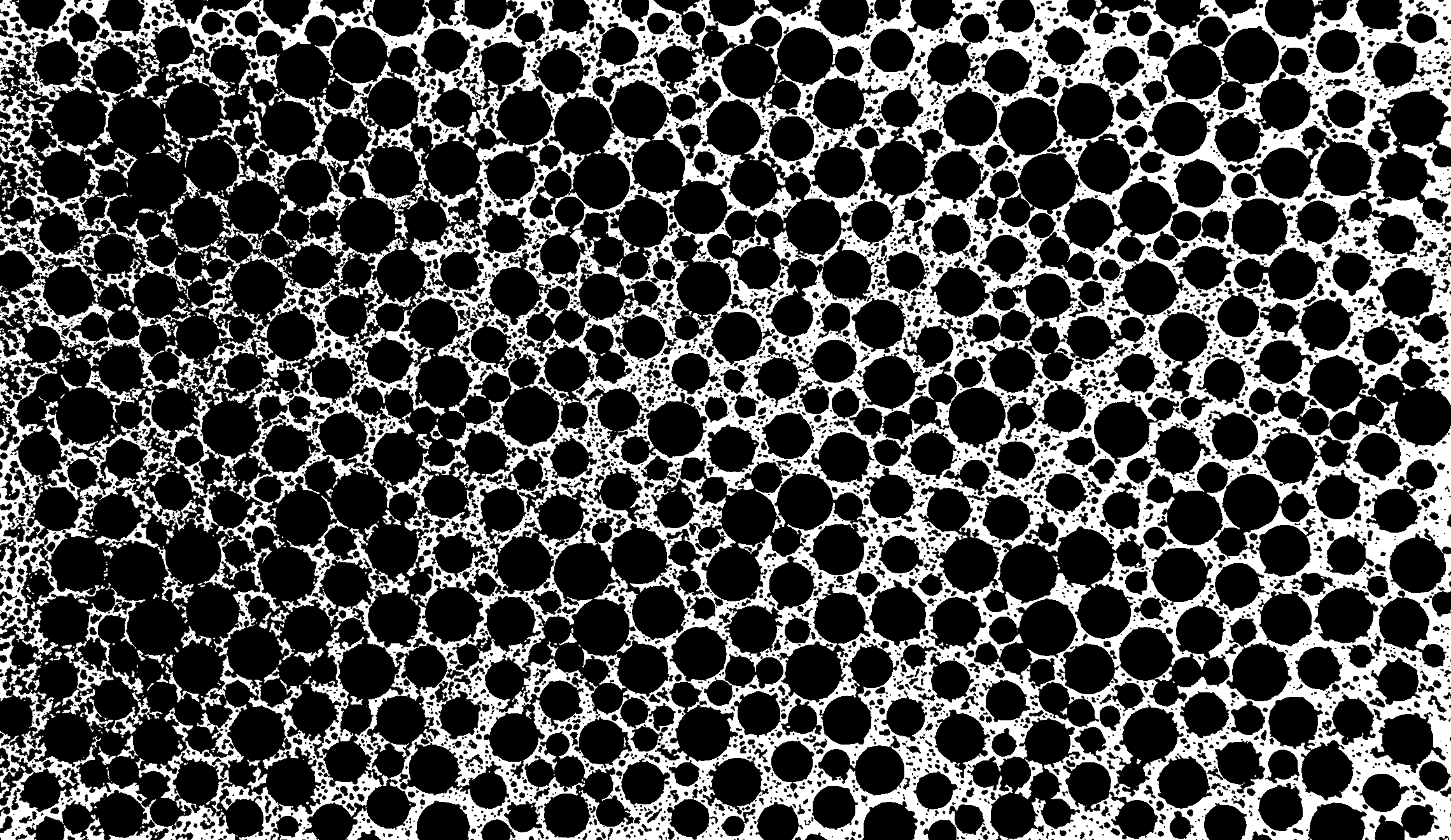}\label{fig:datainit}}
 \subfigure[$t=4$h]{\includegraphics[width=0.33\textwidth]{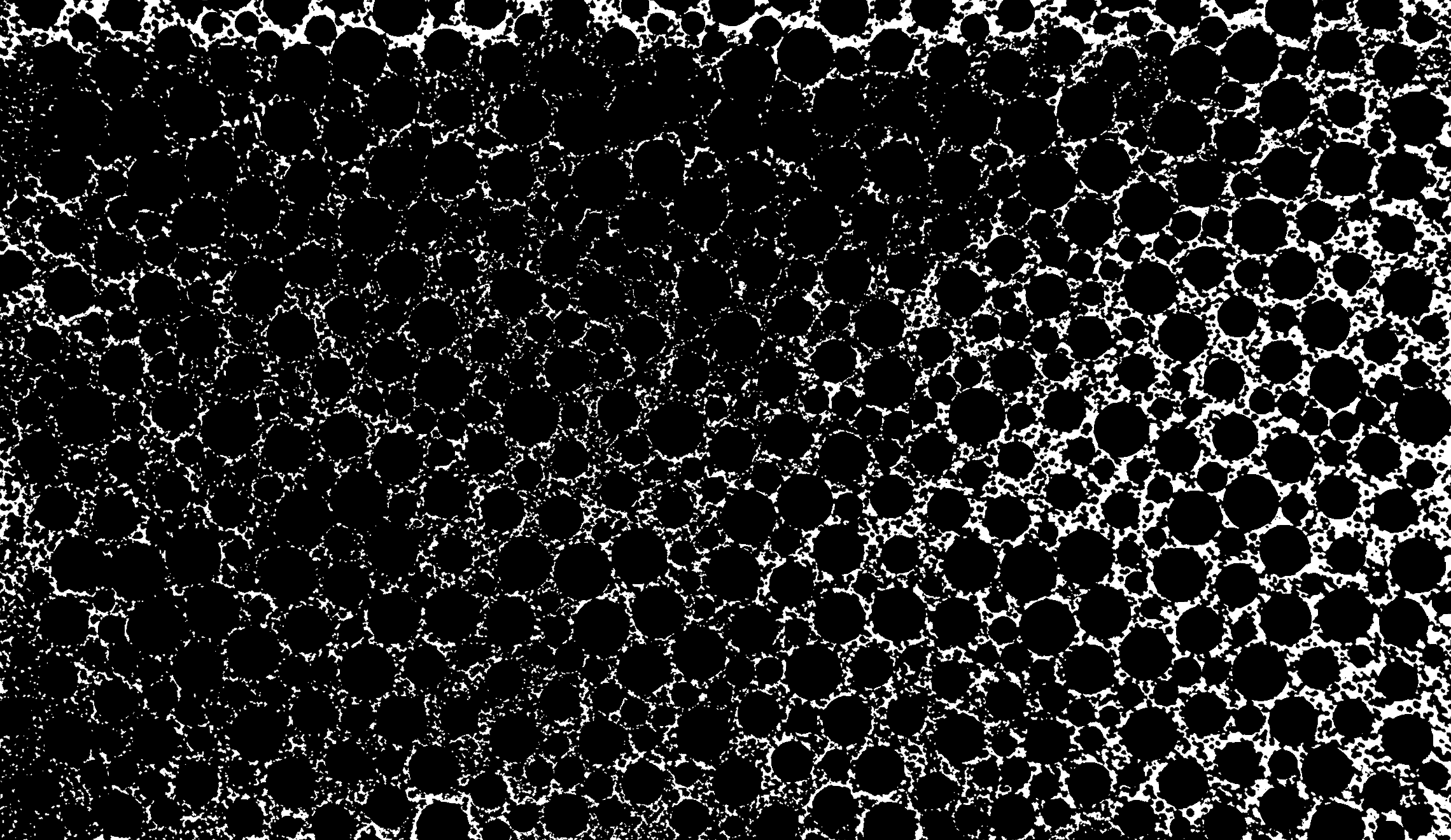}\label{fig:dataend}}\\
 \subfigure[$t=0$h]{\includegraphics[width=0.33\textwidth]{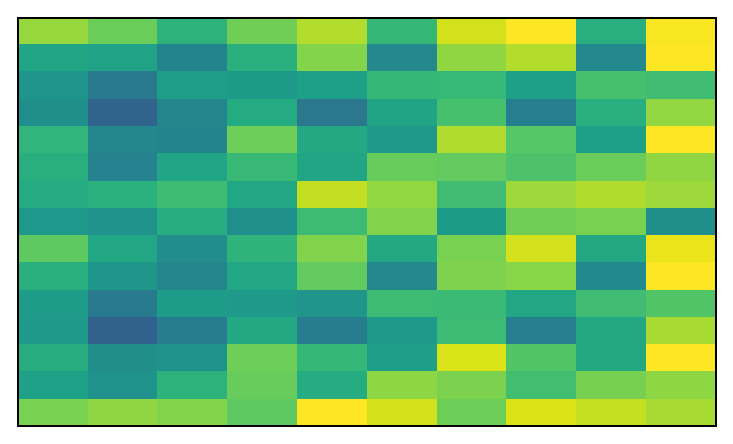}\label{fig:discinit}}
 \subfigure[$t=4$h]{\includegraphics[width=0.33\textwidth]{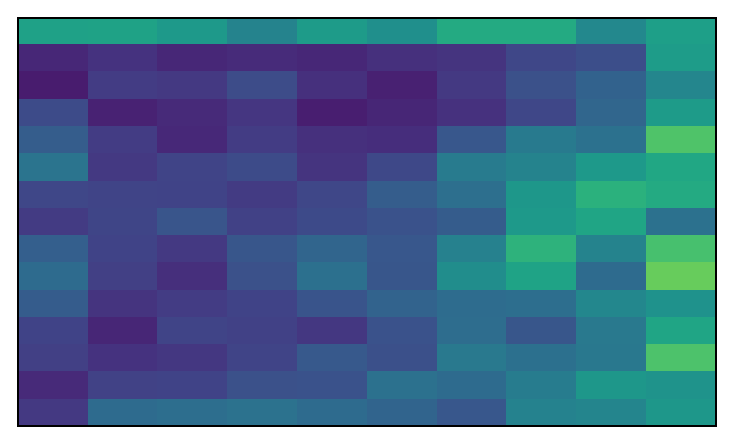}\label{fig:discend}}
\subfigure{\includegraphics[width=0.0445\linewidth]{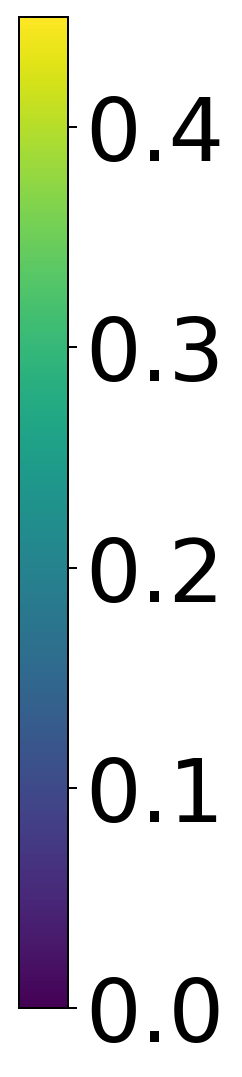}}
\caption{Initial (a) and final (b) segmented images of experiment 'Quasi-2D-1' show appearance of preferential flow at the top of the domain. The discretized porosity fields are shown in (c) and (d). The images were generated with data from \citet{darus-1799_2022}.} \label{fig:data_disc}
\end{figure}
The data of the experiment is given in \citet{darus-1799_2022}, and the initial and final pore geometries are shown in the segmented images in Figure~\ref{fig:data_disc}.
Precipitate volume fractions were obtained as the difference between current and empty cell porosities.
During the experiment, two types of precipitates were observed.
Precipitates that existed as a remainder of the restart appeared darker and were assumed to be calcite or vaterite.
They also did not move from their original positions.
New precipitates appeared first lighter and irregularly shaped, then grew and appeared more dense over time.
This was assumed to be a precursor phase such as ACC.
The lighter precipitates were transported occasionally along with the flow.
Transformation to darker precipitates and subsequent growth was observed in the order of hours.
In this experiment, a visible preferential flow path formed in the upper area of the microfluidic cell.
This flow path stayed open, with increased precipitation in its surrounding.
\citet{weinhardt2022spatiotemporal} hypothesized that due to the transport of small precipitates out of the flow path into its vicinity, the resulting separation of the flow path from the rest of the domain creates a self-enhancing process.

It should be noted that the self-enhancing process of preferential flow path evolution at the top of the microfluidic cell is likely to some extent biased in this experimental setup. The inlet and outlet of the microfluidic cell are positioned in the upper half of the porous domain. Initially, the reactive solutions are expected to flow relatively homogeneously through the porous domain, resulting in a fairly uniform distribution of precipitates. However, as the experiment progresses and a significant amount of precipitates accumulate in the pore space, the fluid increasingly flows through the upper half of the domain where resistance is lower.

\subsection{Model setup}\label{sec:modelDisc}
The two models $M_\mathrm{attached}$ and $M_\mathrm{detached}$ were implemented using the open-source simulator DuMu$^\mathrm{x}$ (DUNE for Multi-{Phase, Component, Scale, Physics, …} flow and transport in porous media) \citep{Koch2021,ScheerClassFlemisch2020}, which is based on DUNE \citep{bastian2008generic1,bastian2008generic2}.
The Euler scheme is used for fully implicit time discretization and the general configuration, except for the spatial discretization, was kept from \citet{hommel2020numerical}.

\begin{figure} [tb]\centering
	\includegraphics[width=0.4\linewidth]{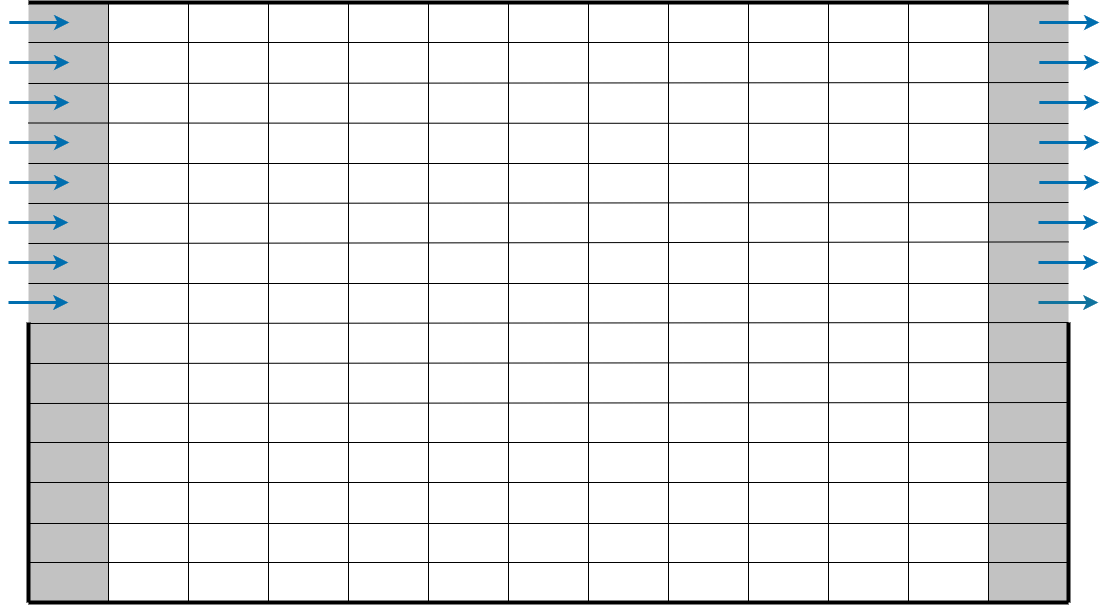}
	\caption{The Experimental domain is extended by two columns (in grey) representing the inlet and outlet distribution channels to allow for the injected fluids distribution over the full height of the microfluidic cell.} \label{fig:modeldomain}
\end{figure}
The discretization scheme was changed from the Box method \citep{ScheerClassFlemisch2020,helmig1997multiphase}, as noted in \citet{hommel2020numerical}, to a cell-centered grid.
The grid-size for the discretization was chosen such that a good resolution of the flow path seen in the experimental data is achieved.
This has the side effect that the columns in the microfluidic cell are in a similar size range as the flow path and, thus, can visibly appear in the discretization.
This grid resolution is too refined for the grid cells to be considered REVs, but necessary to resolve the initial heterogeneity and the evolving flow path.
We choose a discretization that resolves the field into smaller cells in y-direction than x-direction: $10$ cells on the x-axis and $15$ cells on the y-axis.
Figure \ref{fig:data_disc} (c)+(d) visualize the resulting discretizations of the initial and final porosity of the experiment.

To take the inflow and outflow channels into account, the discretized domain was extended by one column of elements on each side as shown in figure \ref{fig:modeldomain}, resulting in a $(15 \times 12)$ grid.
All chemical reactions were disabled in the added cells to avoid an excess of ACC and attached calcite here.
The inlet and outlet are defined over half the height on the left and right side of the modeled domain respectively.

No-flow boundary conditions were set at all edges of the extended domain, except for the inlet.
The inflow was defined via a Neumann boundary condition that matches the applied experimental injection rates and concentrations.
The outlet Dirichlet boundary condition is set equal to the initial condition with urease, ACC, and calcite volume fractions set to zero and the mole fractions of urea, calcium, chloride and urease set according to the concentrations used in \citet{weinhardt2022spatiotemporal}: $0.33\frac{\mathrm{mol}}{\mathrm{l}}$ urea and calcium chloride as well as $2.5\frac{\mathrm{g}}{\mathrm{l}}$ urease.
The model parameter values were taken from the EICP model described in \citet{hommel2020numerical} and extended by the density $\rho_{ACC}=2.18\frac{\mathrm{g}}{\mathrm{cm}^3}$ of ACC\citep{nebel2008kontrollierte,rodriguez2011kinetics}.

\subsection{Surrogate model setup}\label{sec:uncertainparameters}
In order to assess the influence of the parameters introduced by the extension of the model, we need assumptions on their ranges and prior distributions.

\begin{figure}[tbh]
    \centering
	\includegraphics[width=0.65\linewidth]{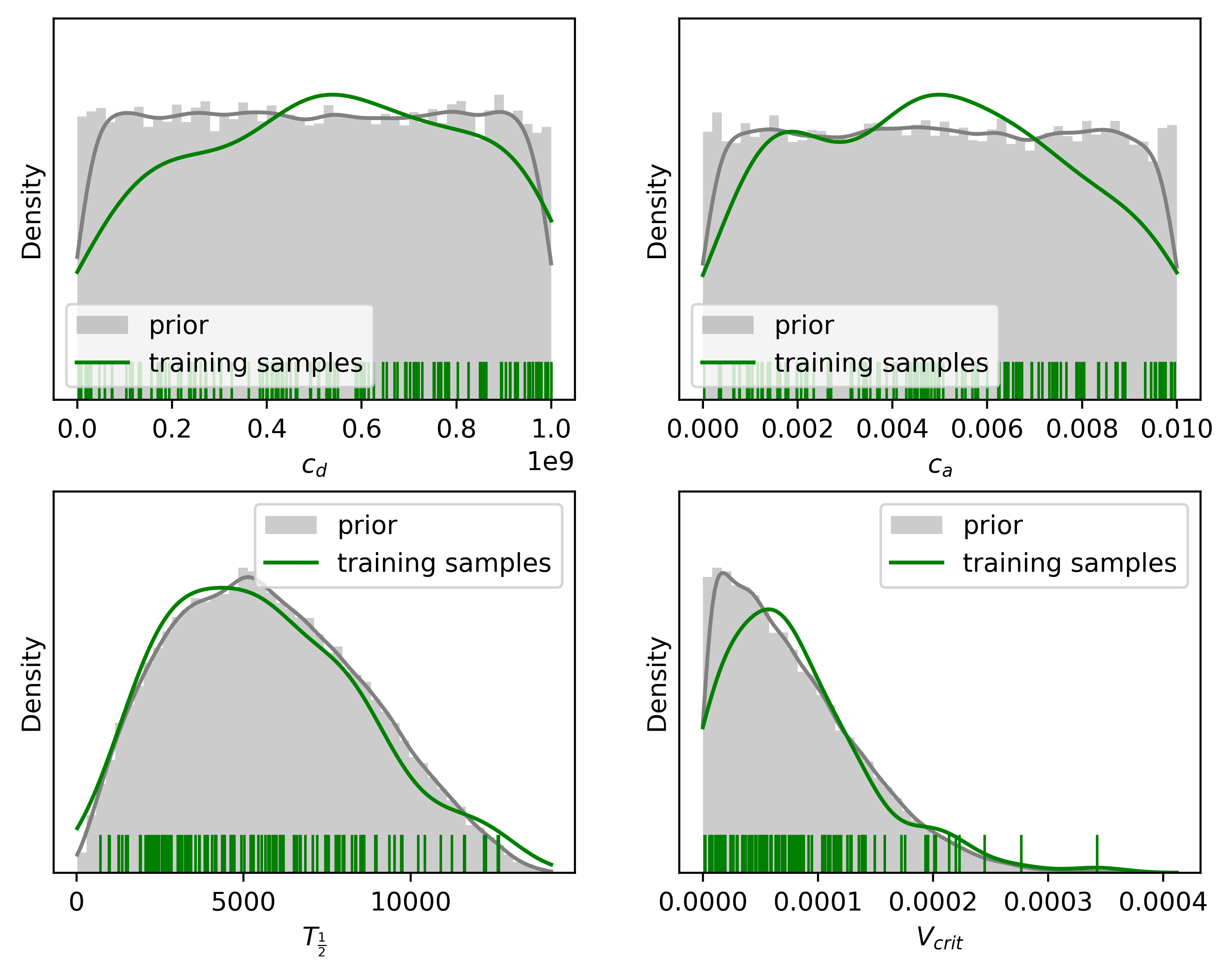}
    \caption{Prior distributions of the unknown input parameters overlaid with the sample distribution from the training set.}
    \label{fig:prior}
\end{figure}

The prior distributions chosen for the four unknown model parameters are given in table \ref{tab:priorDists} and are visualized in figure \ref{fig:prior}.
The priors for the attachment and detachment coefficients are set to uninformative priors in their applicable ranges, which were obtained from test runs of the models.
The half life of ACC was set to a scaled Beta distribution and cut off at the experiment length of four hours.
For the critical velocity, a lower bound of $0$ follows from its definition, but no upper bound could be given.
These distributions are used to choose evaluation points and perform parameter estimation.

\begin{table}[tb]
    \centering
    \begin{tabular}{p{0.15\textwidth}|p{0.14\textwidth}|p{0.16\textwidth}}
        \toprule
        Parameter  & Bounds & Distribution\\
       \midrule
         $c_\mathrm{a}$ & $[0,0.003]$ & $\mathcal{U}(0,0.003)$\\
         $c_\mathrm{d}$ &  $[0,10^6]$ &
          $\mathcal{U}(0,10^6)$\\
         $T_\frac{1}{2}$ &  $[0,14400]$ & $\mathcal{B}(2,3)\cdot 14400$\\
         $v_\mathrm{Crit}$ &  $[0,\infty)$ & $\mathcal{N}(0,10^{-4})$\\
        \bottomrule
    \end{tabular}
    \caption{Prior distributions for the model parameters.}
    \label{tab:priorDists}
\end{table}

The surrogate models, Bayesian inference and model comparison are implemented with BayesValidRox \citep{darus-4613_2024}, a python package that contains Bayesian validation methods for computational models.
The surrogate models for each of the variations $M_\mathrm{attached}$ and $M_\mathrm{detached}$ were trained on $128$ model evaluations on randomly chosen samples of the parameter prior distributions. 
The distributions of the training samples are overlayed in green in figure \ref{fig:prior}.

For the Bayesian model analysis described in section \ref{sec:Bayes}, we assume the uncertainty of the porosity values to follow a Gaussian distribution $\mathcal{N}(\phi,\sigma(\phi))$ around the measured porosity values $\phi$.
The uncertainty of each evaluation is estimated as a combination of the discretization error from the image pixel size, approximation errors from assuming a frustum shape for the precipitates, and numerical errors that stem from simplifications during the porosity calculation.
Other possible error sources include possible cut off of the microfluidic cell at the image domain and the inherent uncertainty of the optical microscopy results.

\section{Results and discussion}\label{sec:results}
In this section we present the modeling and evaluation results on the setup described in section \ref{sec:setup}.

\subsection{General model behavior}
\begin{figure}[tb]\centering
	\subfigure[]{\includegraphics[width=0.2\linewidth]{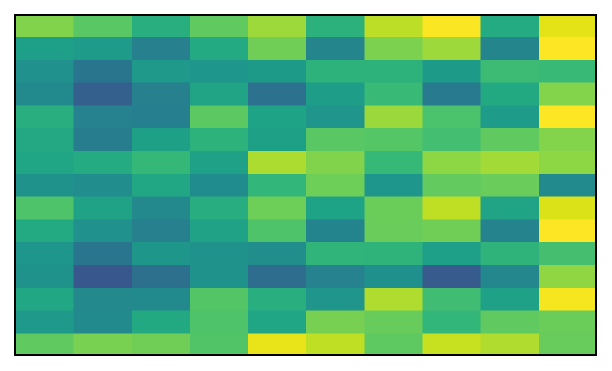}\label{fig:paths_a}}
	\subfigure[]{\includegraphics[width=0.2\linewidth]{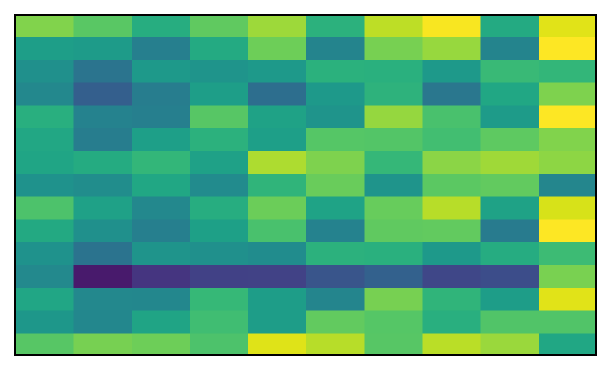}\label{fig:paths_b}}
	\subfigure[]{\includegraphics[width=0.2\linewidth]{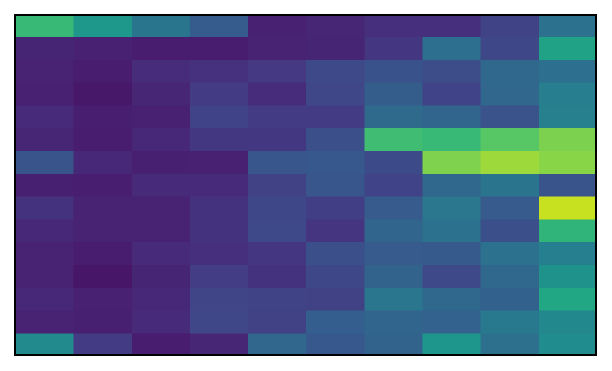}\label{fig:paths_c}}
	\subfigure[]{\includegraphics[width=0.2\linewidth]{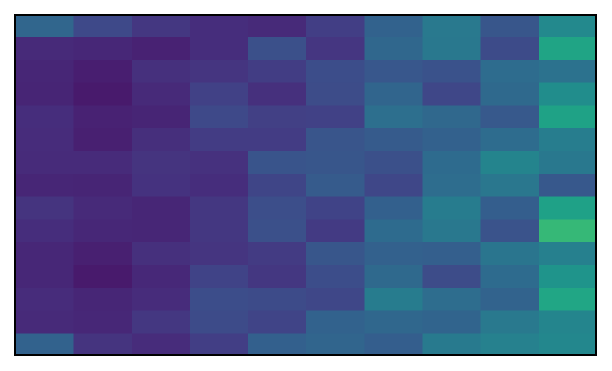}\label{fig:paths_d}}
 
	\subfigure[]{\includegraphics[width=0.2\linewidth]{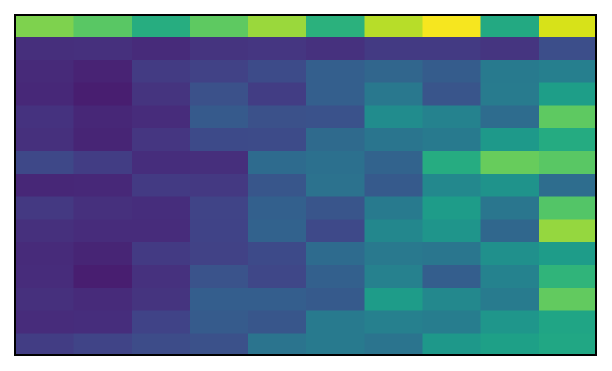}\label{fig:paths_e}}
	\subfigure[]{\includegraphics[width=0.2\linewidth]{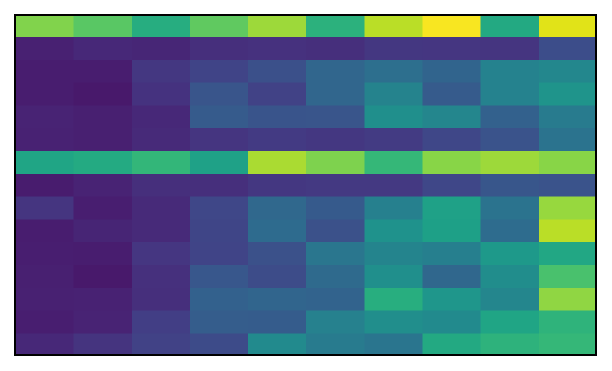}\label{fig:paths_f}}
	\subfigure[]{\includegraphics[width=0.2\linewidth]{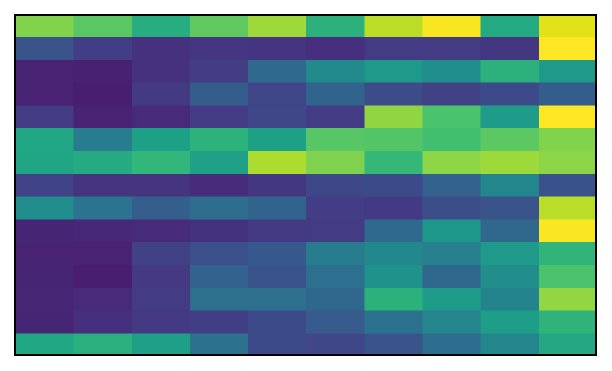}\label{fig:paths_g}}
	\subfigure[]{\includegraphics[width=0.2\linewidth]{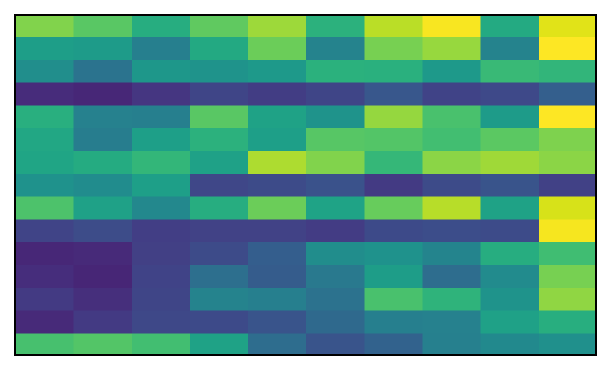}\label{fig:paths_h}}
 
	\subfigure[]{\includegraphics[width=0.2\linewidth]{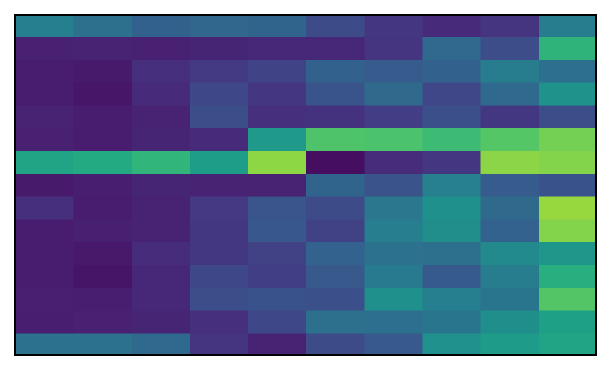}\label{fig:paths_i}}
	\subfigure[]{\includegraphics[width=0.2\linewidth]{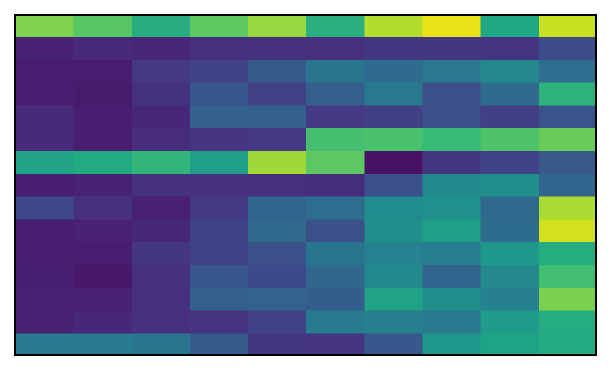}\label{fig:paths_j}}
	\subfigure[]{\includegraphics[width=0.2\linewidth]{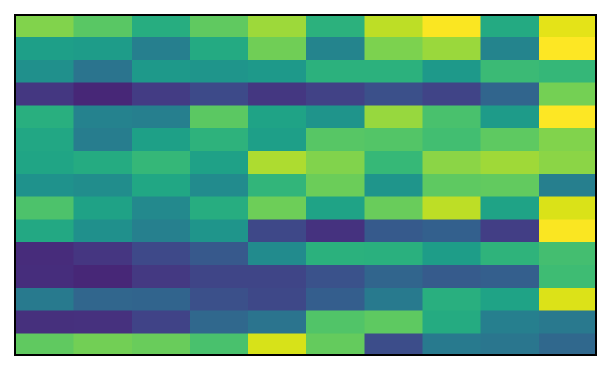}\label{fig:paths_k}}
    \subfigure{\includegraphics[width=0.03\linewidth]{Fig7_l.png}}
    \caption{Porosity of model evaluations at $t=4$h show flow paths of different shapes, predominantly linear at top and middle of the domain.}
    \label{fig:modelflowpaths}
\end{figure}

We consider the porosity values in all discrete spatial volumes at $t=4$h to be the primary results of each model evaluation.
Representative porosity fields for different model evaluations are collected in figure \ref{fig:modelflowpaths}, with darker values indicating lower porosity.
The model results vary visibly in their amounts of precipitation and show appearing flow paths of different intensities and shapes.
The first row of the figure contains model results with a mostly homogeneous porosity.
These results indicate that for the parameter priors defined in section \ref{sec:uncertainparameters}, we can span the full range of attachment to detachment ratios, from predominant detachment for \ref{fig:paths_a} and \ref{fig:paths_b} to predominant attachment for \ref{fig:paths_c} and \ref{fig:paths_d}.
For cases in between these two extremes, we observe the formation of preferential flow paths, characterized by a series of connected, high-porosity volumes.
The most common shape of preferential flow paths found in the model evaluations are horizontal linear flow paths at the top and the middle of the domain, with occasionally additional flow path at the bottom of the domain, see \ref{fig:paths_c}-\ref{fig:paths_h}.
The linear flow paths themselves can vary in height and intensity and are bordered by cells with lower porosity.
Here, we want to highlight figure \ref{fig:paths_e}, which shows a single linear flow path at the top of the domain, similarly to the final state of the experiment, see \ref{fig:discend}.
Curved flow paths also appear for both model variations, mostly at the bottom (see \ref{fig:paths_k}) and middle of the domain (see \ref{fig:paths_i}-\ref{fig:paths_j}).
They appear both alone and in combination with other flow paths.

\begin{figure}[tb]\centering
	\subfigure[Porosity]{\includegraphics[width=0.3\linewidth]{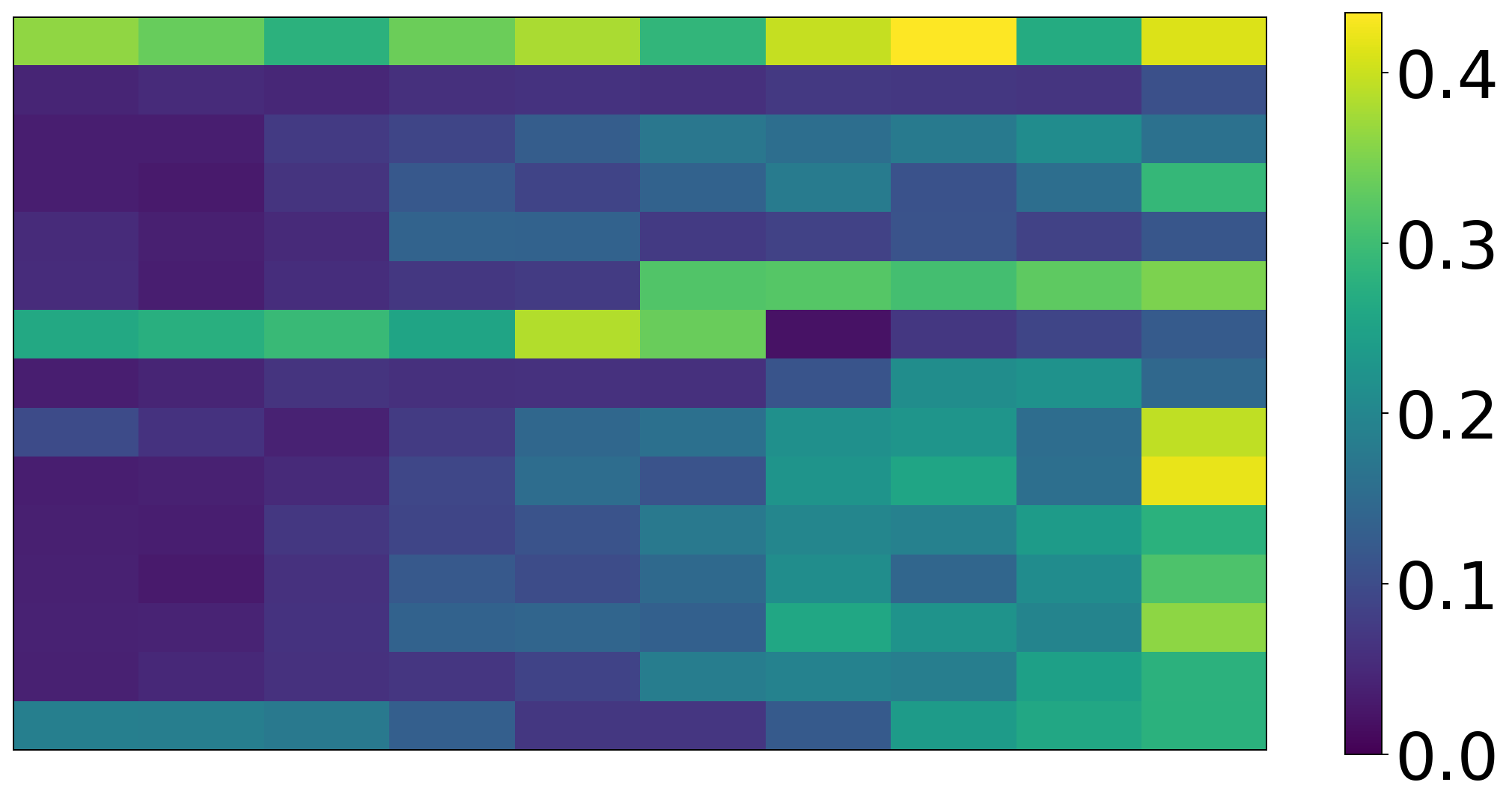}\label{fig:model_poro}}
	\subfigure[Velocity]{\includegraphics[width=0.33\linewidth]{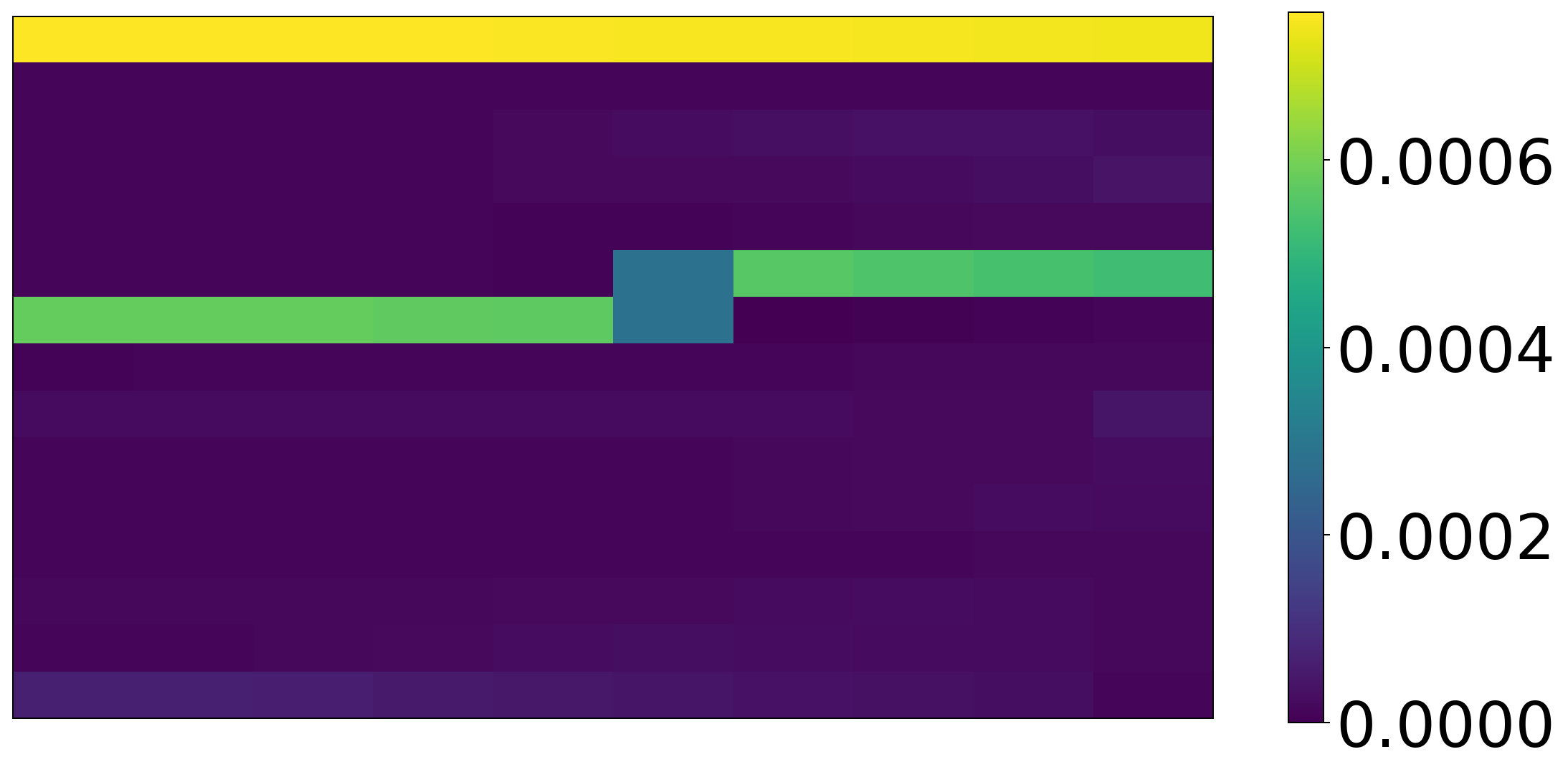}\label{fig:model_vel}}
	\subfigure[Calcite]{\includegraphics[width=0.3\linewidth]{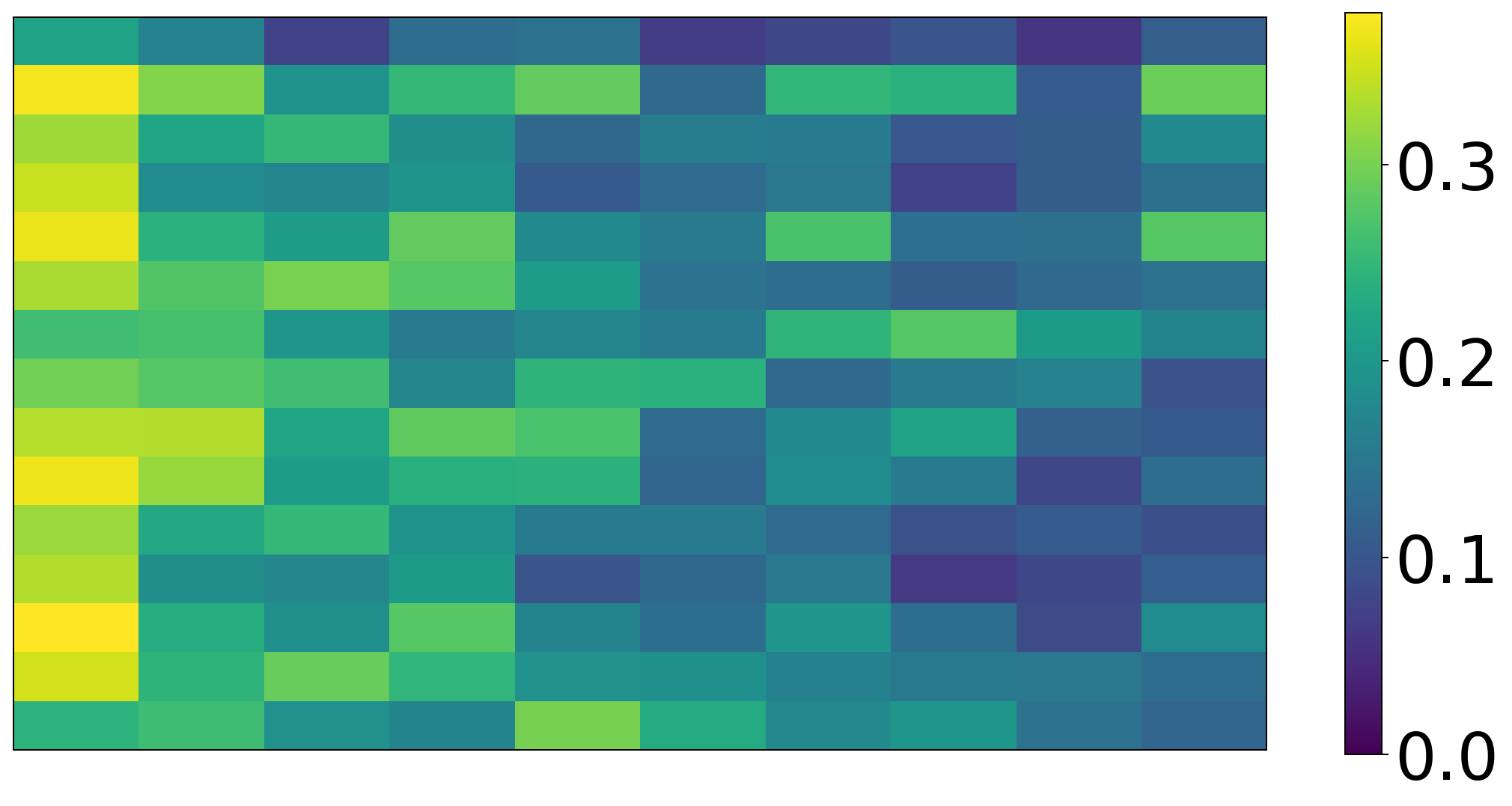}\label{fig:model_calc}}
	\subfigure[Detached ACC]{\includegraphics[width=0.32\linewidth]{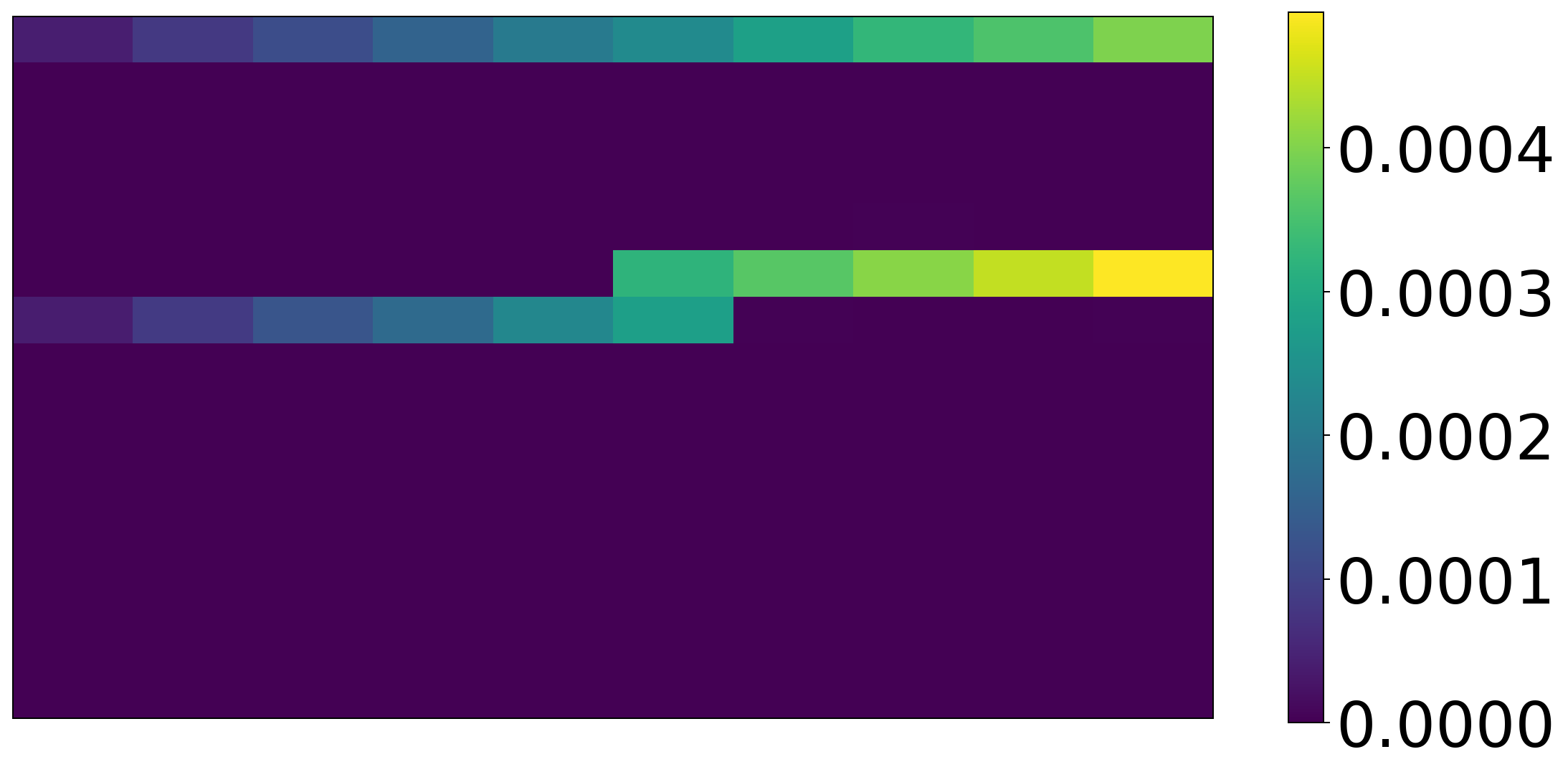}\label{fig:model_ACCdet}}
	\subfigure[Attached ACC]{\includegraphics[width=0.3\linewidth]{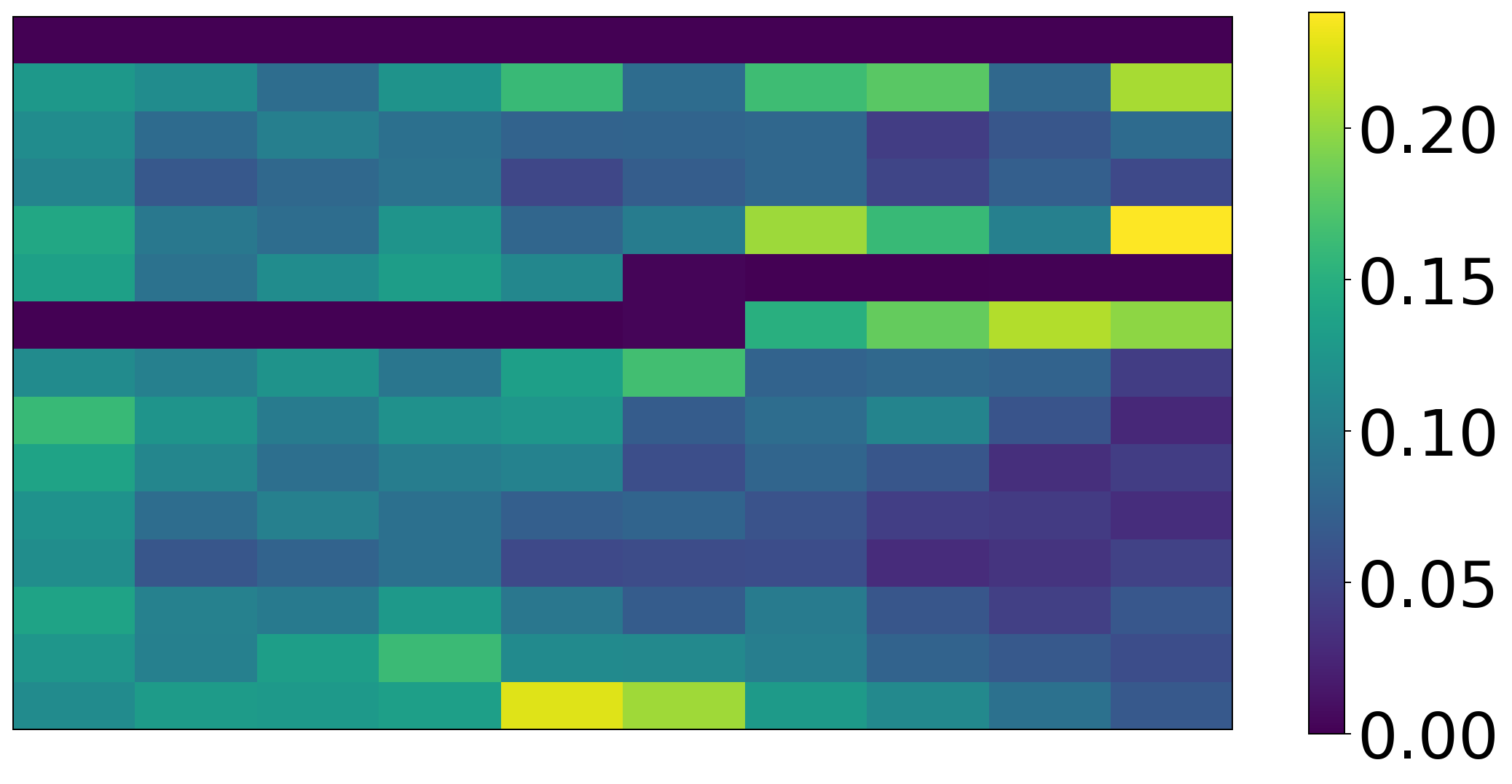}\label{fig:model_ACCatt}}
	\caption{Final state of the model quantities porosity, velocity, calcite and detached and attached ACC at $t=4$h for a representative model evaluation.}
    \label{fig:modelfields}
\end{figure}

The modeled porosity depends on both the precipitate volume fractions of attached ACC $\phi_\mathrm{ACC}$ and calcite $\phi_\mathrm{c}$, see Eq.~\eqref{eq:porosity}. 
Further, high velocities $v$ exceeding the critical velocity, as found in flow paths, lead to detachment of ACC, see Eq.~\eqref{eq:ACCdetach}.
Thus, we expect to find most calcite and attached ACC outside of flow paths, where the porosities are low.
In contrast, detached ACC and high velocities are expected within flow paths, where the porosities are high.
This effect is visualized with the results of a representative model evaluation in figure \ref{fig:modelfields}.
The model evaluation produced two dominating flow paths, one at the very top and one in the middle of the domain.
We observe that the velocities in figure~\ref{fig:model_vel} scale mainly with the height of the flow paths, while the concentration of detached ACC in figure~\ref{fig:model_ACCdet} increases over the length of the flow path.
The precipitated calcite, see figure \ref{fig:model_calc}, forms a gradient from left to right with visibly reduced amounts of calcite in the flow paths.
In comparison, the attached ACC, see figure \ref{fig:model_ACCatt}, can be found mainly at the edges of flow paths, where the velocities are lower and the porosity not too small.

\subsection{Influence of the initial porosities}\label{sec:inletporo}
\begin{figure}[tbh]\centering
	\subfigure[Full homogeneous porosity]{\includegraphics[width=0.31\linewidth]{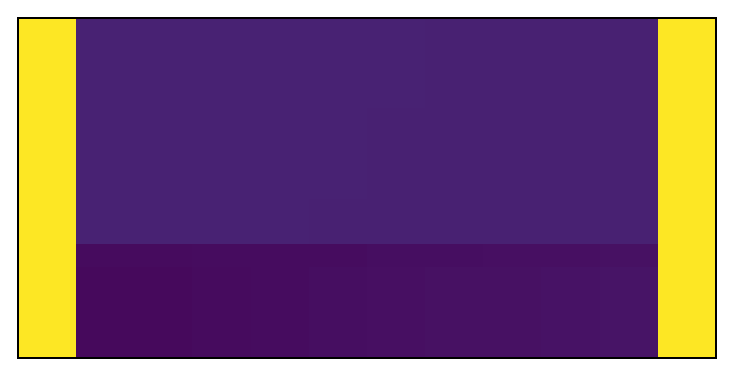}\label{fig:poro_homogeneous}}
	\subfigure[Homogeneous inlet]{\includegraphics[width=0.31\linewidth]{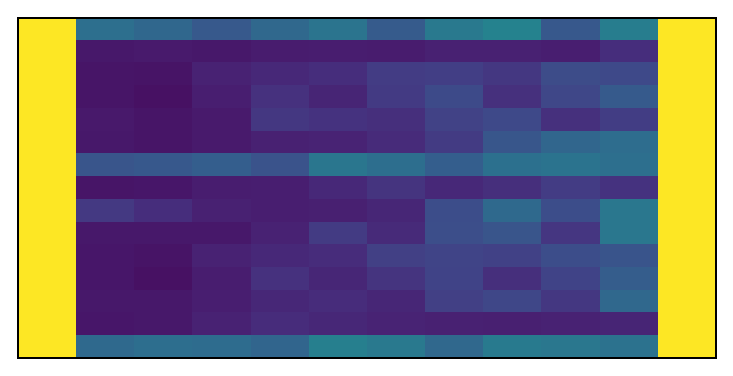}\label{fig:poro_homogeneousinlet}}
	\subfigure[Inhomogeneous inlet]{\includegraphics[width=0.31\linewidth]{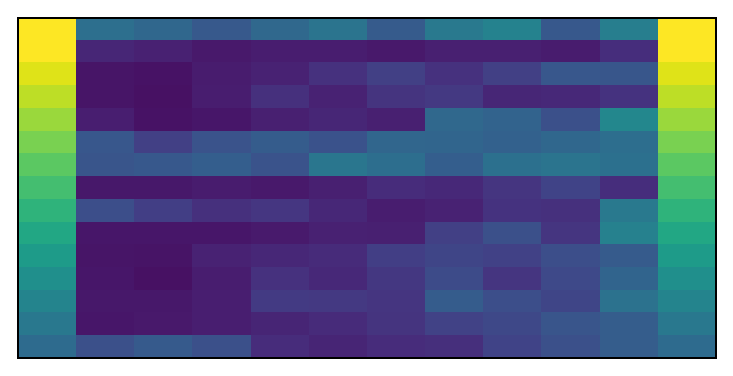}\label{fig:poro_heterogeneous}}
    \subfigure{\includegraphics[width=0.038\linewidth]{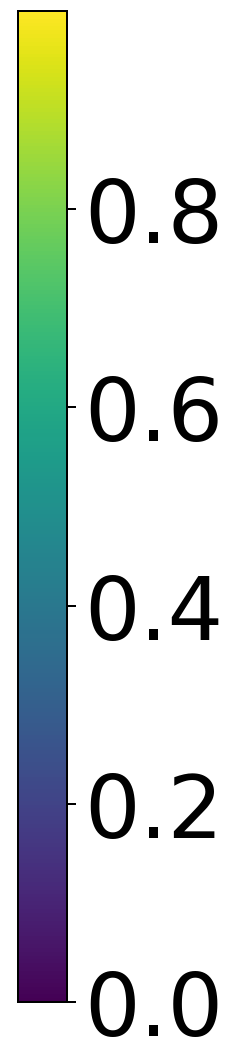}}
    \caption{Final full porosity fields for the same model parameter values, but different inlet porosities.}
    \label{fig:porosityInfluence}
\end{figure}
Based on the setup described in section \ref{sec:model}, we set two types of initial porosities for the model, i.e., the porosity in the microfluidic cell and the porosity in the extensions for the inlet and outlet.
We expect both of these to impact the formation and placement of preferential flow paths in the model evaluations.

Figure \ref{fig:porosityInfluence} shows the final porosities for model runs with identical parameter values, but different initial porosity distributions.
The evaluation in figure \ref{fig:poro_homogeneous} uses a homogeneous initial porosity of $\phi=0.1$ everywhere except the in- and outlet and no initial calcite as well as homogeneous initial surface area.
In contrast, the model evaluations for figure \ref{fig:poro_homogeneousinlet} and figure \ref{fig:poro_heterogeneous} use the heterogeneous porosity, initial calcite, and initial surface area that were estimated from the experiment setup in section \ref{sec:experiment}.
The porosity field for the fully homogeneous setup remains homogeneous and without flow paths, while the heterogeneous setups result in multiple clear flow paths.
This supports the assumption described in section~\ref{sec:intro}, that initial heterogeneity is required to start the formation of preferential flow paths.
The two heterogeneous setups in figure \ref{fig:porosityInfluence} differ only in the initial porosity for the inlet and outlet.
We set homogeneous porosity for the in- and outlet in figure \ref{fig:poro_homogeneousinlet} and reduce the porosity towards the bottom of the domain for the in- and outlet in figure \ref{fig:poro_heterogeneous} to better match the experimental setup.
The homogeneous inlet and outlet porosities lead to overall much clearer defined straight flow paths at the top, middle, and bottom of the domain.
Decreasing in- and outlet porosity towards the bottom of the domain reduces the intensity of the flow path at the bottom, and the middle flow path is strengthened.

For all model evaluations in this work we use the initial porosities as described for figure \ref{fig:poro_heterogeneous}, with heterogeneous initial porosity in the model domain and reduced porosity towards the bottom of the in- and outlet.
We set these initial porosities to obtain flow paths from the models that are more comparable to the experiment, without attempting an exact replication of the experimental results.

\subsection{Parameter behavior}\label{sec:parameters}
We quantify the dependency of the model evaluations on the four uncertain input parameters presented in section~\ref{sec:uncertainparameters} with the help of Sobol' indices as described in section~\ref{ssec:sobol}.
This section presents the overall behavior of the extended model based on results for $M_\mathrm{attached}$, the comparison between $M_\mathrm{attached}$ and $M_\mathrm{detached}$ is performed in section \ref{sec:modelcomp}.
Due to the structure of the built surrogate models, we obtain a set of Sobol' indices and Total Sobol' indices for each discrete point in time and space.

\begin{figure}[tb]\centering
	\subfigure{\includegraphics[width=0.45\textwidth]{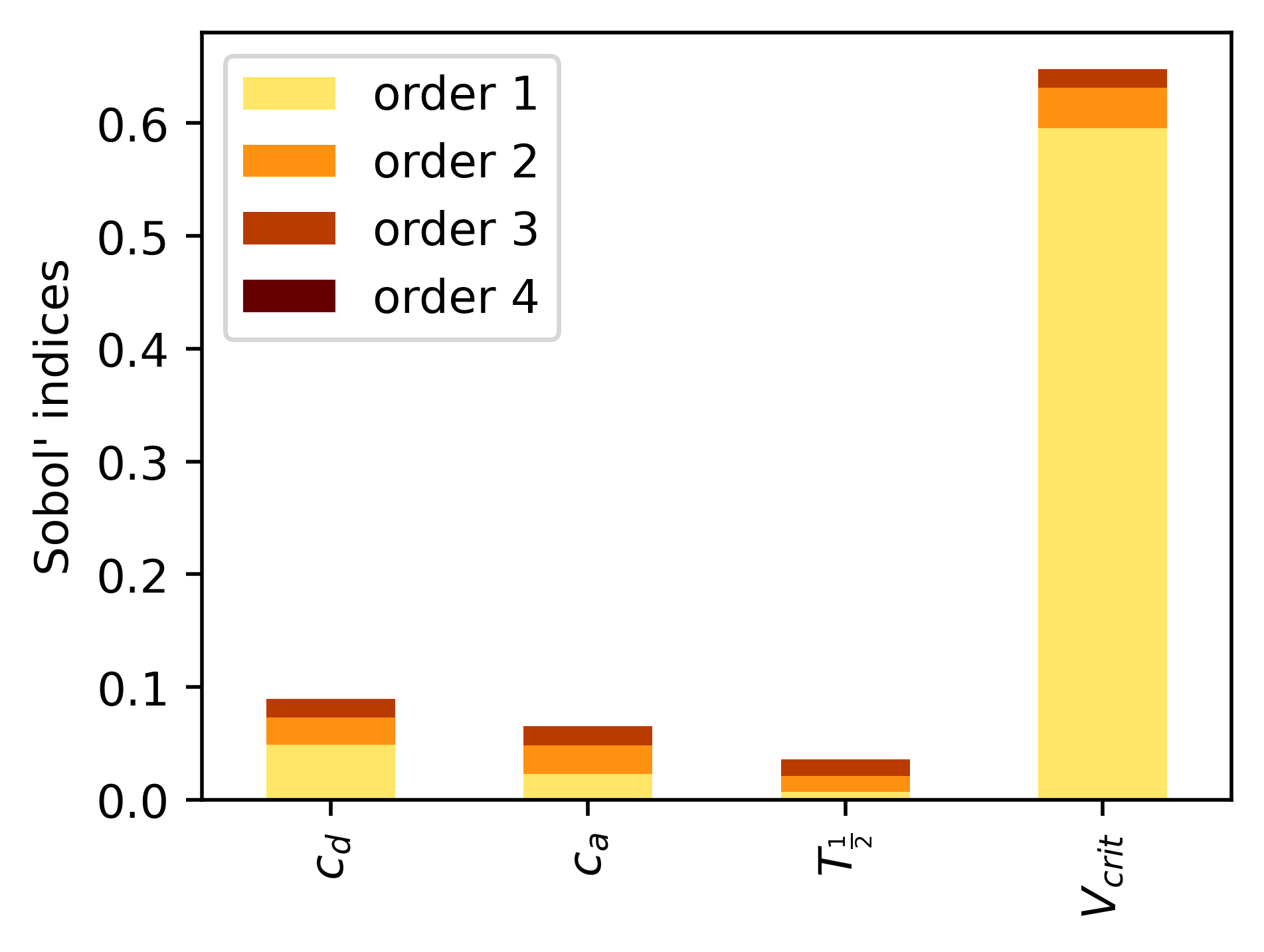}}
    \caption{Sobol' indices for model $M_{attached}$ averaged over all points in space and time.}
    \label{fig:sobol_avg}
\end{figure}
We first consider the Sobol' indices averaged over all points in space and time to quantify the overall importance of each parameter. 
The four unknown parameters $c_\mathrm{a}$, $c_\mathrm{d}$, $T_{\frac{1}{2}}$ and $v_\mathrm{Crit}$ of the extended EICP model are involved in processes that are inherently strongly coupled.
Thus, we consider not only the first order Sobol' indices, that describe the individual contribution of each parameter, but also higher order Sobol' indices that describe the joint sensitivity of the model on sets of two or more parameters.
Figure \ref{fig:sobol_avg} visualizes the averaged Sobol' indices of orders $1$ to $4$ for each input parameter.
As there are four unknown parameters, indices of higher orders than $4$ do not appear.
The figure shows that the model is most sensitive to the critical velocity $v_\mathrm{Crit}$, with both the largest first order Sobol' index at approximately $0.6$ and the largest overall contribution at approximately $0.65$.
The critical velocity acts as a threshold value that needs to be surpassed for detachment to occur.
It scales not only the detachment, but also influences the attachment processes, which depend on the amount of detached ACC in the system, and in turn are a prerequisite for the transformation to calcite.
The high importance of the critical velocity parameter means that in comparative settings this parameter is most important to describe how the model behaves.
The other three parameters each have a total sensitivity of less than $0.1$.
The sensitivities of the attachment and detachment coefficients $c_\mathrm{a}$ and $c_\mathrm{d}$ are similar in split between individual and combined contributions.
The slightly larger Sobol' indices for $c_\mathrm{d}$ is explained due to its direct coupling with the critical velocity.
Of the four parameters, the half life $T_{\frac{1}{2}}$ of ACC is the least important, with almost no individual contribution.

\begin{figure}[tb]\centering
	\subfigure[Total' Sobol indices at $(4,7)$]{\includegraphics[width=0.32\linewidth]{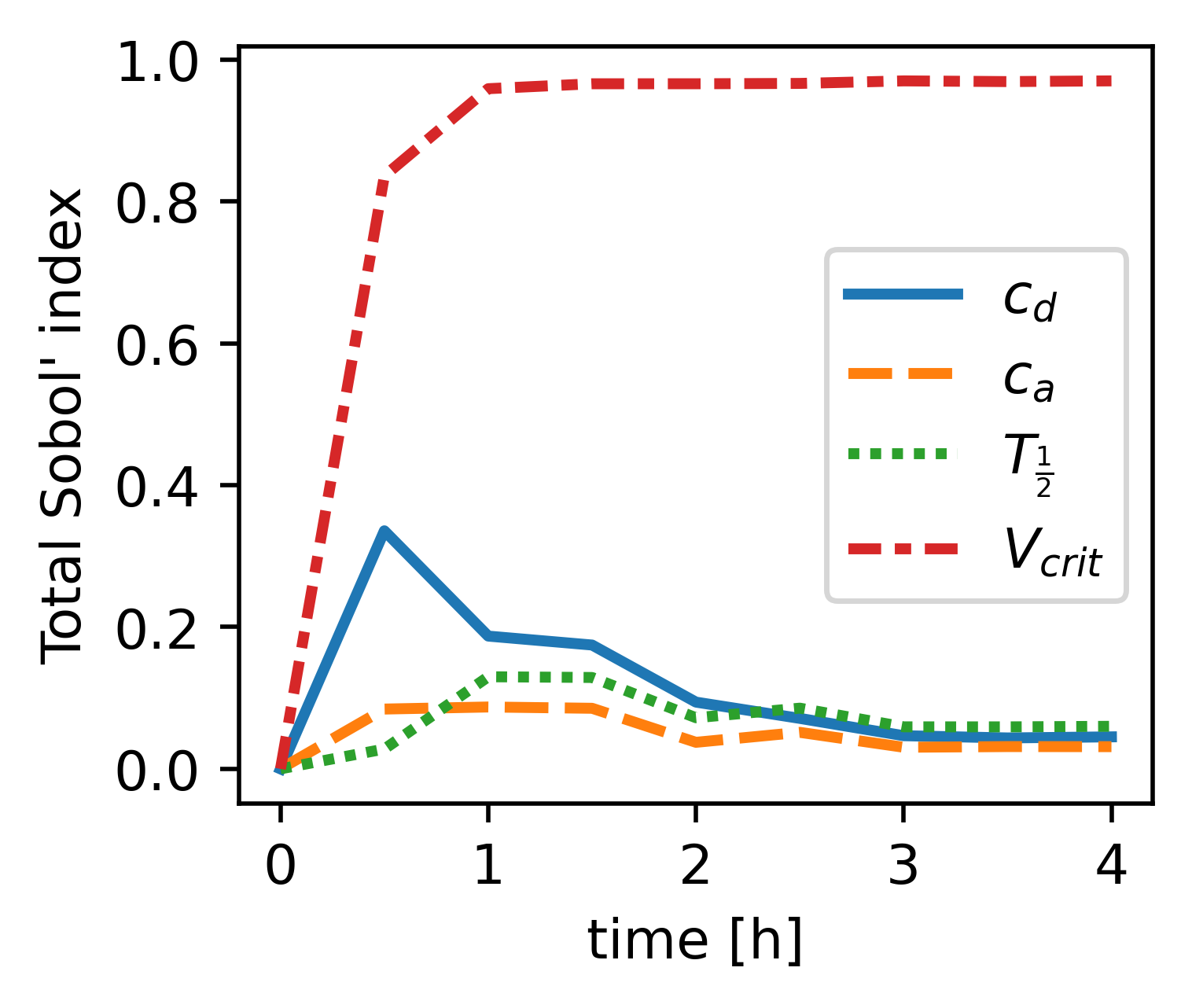}\label{fig:sobol_flowpath}}
	\subfigure[Total' Sobol indices at $(10,5)$]{\includegraphics[width=0.32\linewidth]{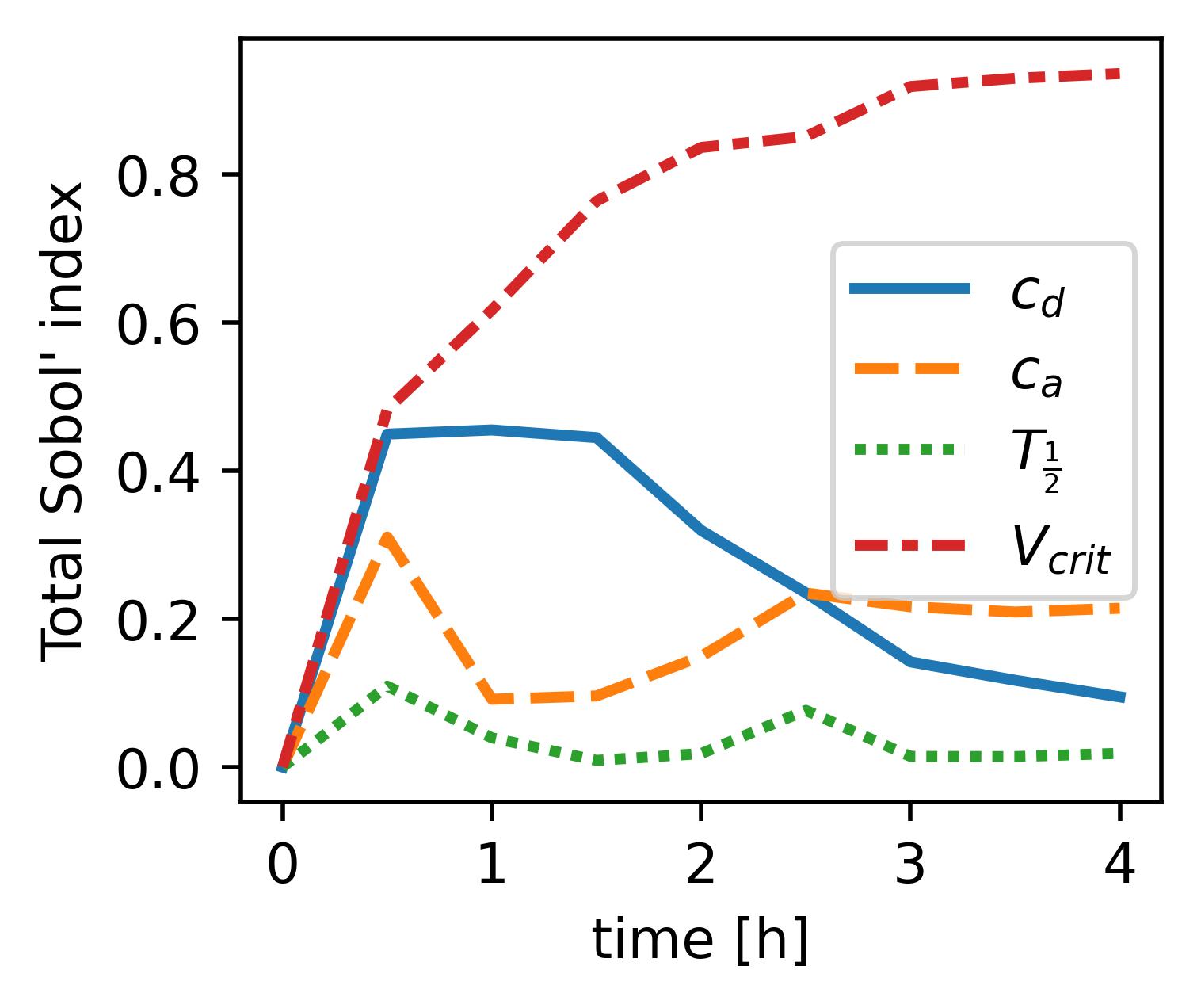}\label{fig:sobol_mixed}}
	\subfigure[Total' Sobol indices at $(2,12)$]{\includegraphics[width=0.32\linewidth]{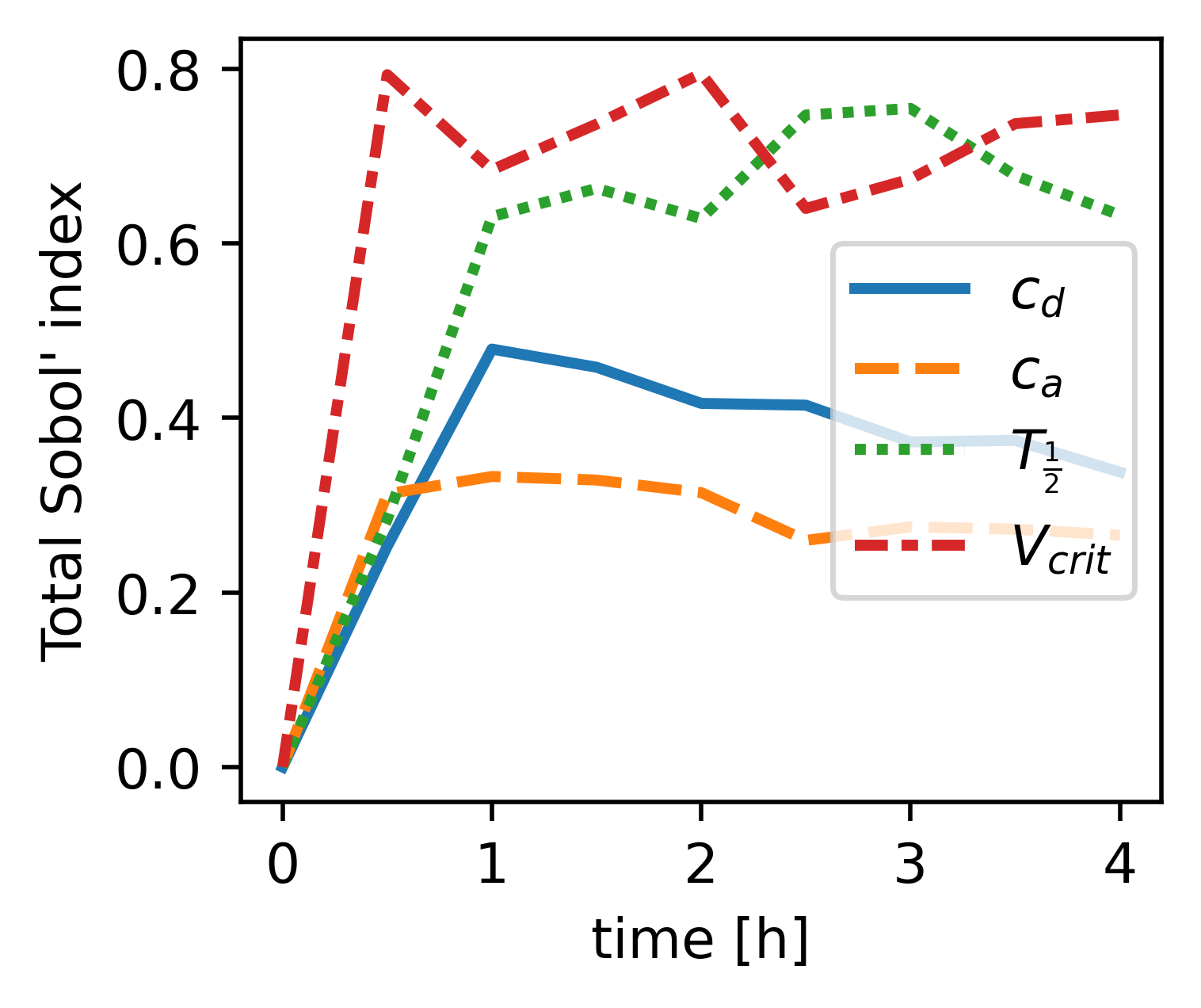}\label{fig:sobol_noflow}}
	\subfigure[Averaged porosity field at $t=4$h]{\includegraphics[width=0.45\textwidth]{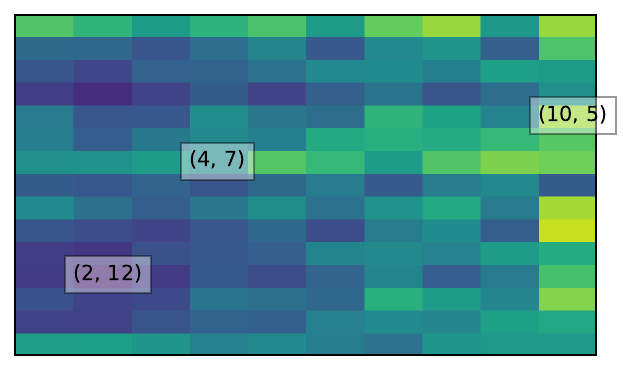}\label{fig:poro_avg}}\vfill
    \caption{Total Sobol' indices for three spatial cells of model $M_{attached}$ show differences in dependence on the uncertain parameters.}
    \label{fig:sobolM_attached}
\end{figure}
The high importance of the critical velocity can be further explained from sensitivity analysis for specific points in space.
Figure \ref{fig:sobolM_attached} shows the Total Sobol' indices over time for three different spatial cells.
The cells were chosen based on how often they are part of formed flow paths in the model evaluations used as training data.
Figure \ref{fig:poro_avg} overlays the location of the three chosen cells on the average porosity field of all model evaluations at $t=4$h.
Cell $(4,7)$ lies primarily in formed flow paths, while cell $(10,5)$ shows mixed behavior.
Cell $(2,12)$ represents areas that are unlikely to join flow paths.

The Total Sobol' indices start at $0$ for $t=0$~h for all cells, as no sensitivity can be calculated while no changes have occurred.
The evolution of cells that are highly likely to be part of a flow path, shown in figure~\ref{fig:sobol_flowpath}, is dominated by the critical velocity.
Whether such a cell becomes part of a flow path is not decided by the interplay of attachment and detachment, but by the velocity in the cell.
There is a slight sensitivity to the other three parameters, but it decreases to $0$ over time.
In contrast, in cells that are highly unlikely to be part of a formed flow path, the sensitivity with respect to the four parameters is visibly more balanced, as shown in figure~\ref{fig:sobol_noflow}.
Here, the half life $T_{\frac{1}{2}}$ of ACC is of similar importance as the critical velocity, with the coefficients that scale the attachment and detachment not far behind.
Cells that show a mixed behavior, see figure~\ref{fig:sobol_mixed}, start out without a clearly dominating parameter.
Over time, as their path towards being part of a flow path or not is decided, they transition to a similar sensitivity as described for figure~\ref{fig:sobol_flowpath}, with high dependency on the critical velocity.
In general, we conclude for this model that the more the evolution of a spatial cell is dominated by the critical velocity $v_\mathrm{Crit}$, the more likely it is to be part of a formed flow path.

\subsection{Model comparison}\label{sec:modelcomp}
We compare the model variants $M_\mathrm{attached}$ and $M_\mathrm{detached}$ that were introduced in section \ref{sec:extendModel} based on three aspects, i.e., representative model outputs for different parameter combinations, the sensitivity of the models on the input parameters, and Bayesian model analysis as described in section \ref{sec:MJA}.

\begin{figure}[tb]\centering
	\subfigure[]{\includegraphics[width=0.3\linewidth]{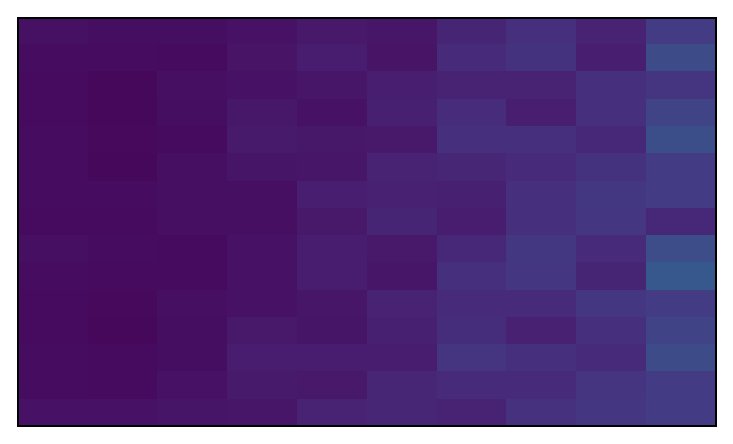}}
	\subfigure[]{\includegraphics[width=0.3\linewidth]{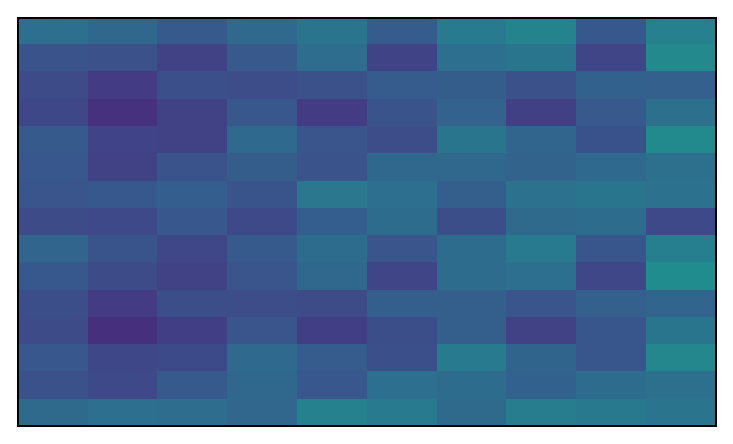}}
    \subfigure{\includegraphics[width=0.04\linewidth]{Fig7_l.png}}
    \caption{Model outputs of Low $c_\mathrm{a}$ and $c_\mathrm{d}$ lead to expected difference in model evaluations at $t=4$~h between $M_\mathrm{attached}$ and $M_\mathrm{detached}$.}
    \label{fig:modelcomp_sample}
\end{figure}

We expect that for the same model parameters, $M_\mathrm{attached}$ will result in higher amounts of attached precipitate and, accordingly, smaller final porosities than $M_\mathrm{detached}$.
An example is shown in figure \ref{fig:modelcomp_sample}.
Here, we consider the extreme case and set $c_\mathrm{a}=0$ and $c_\mathrm{d}=0$, which allows no attachment or detachment after the initial precipitation of ACC.
Consequently, $M_\mathrm{detached}$ keeps porosity constant from the initial to the final state of the model, while $M_\mathrm{attached}$ holds all precipitated ACC in the domain.
For non-zero values of $c_\mathrm{a}$ and $c_\mathrm{d}$ we observe a 'self-regulatory' effect of the attachment and detachment processes, where the processes find a dynamic equilibrium independent of the underlying assumption about the initial ACC state.
To have high attachment in a time step, a large amount of previously detached ACC is required in the system, and vice versa.
When comparing $M_\mathrm{attached}$ and $M_\mathrm{detached}$, similar equilibrium states seem to be reached, resulting in almost identical porosity distributions for both models across a broad range of the investigated parameter ranges of $c_\mathrm{a}$ and $c_\mathrm{d}$.
Since this effect is present in most of the obtained model evaluations, it dominates the comparison of the model variations.
As a result, the Sobol' indices of the two models are almost identical, with only slight shifts of the importance of $c_\mathrm{a}$ and $c_\mathrm{d}$.

\begin{figure}[tb]\centering
	\subfigure[Confusion matrix]{\includegraphics[width=0.5\linewidth]{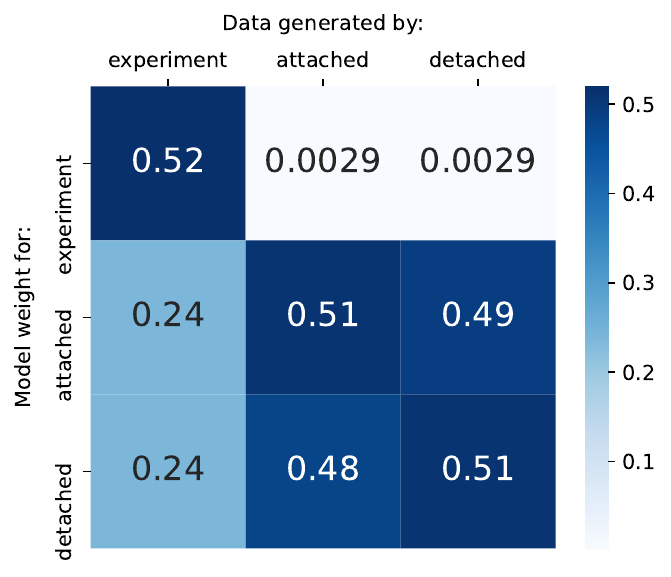}}
    \caption{Confusion matrix for the experimental data and models $M_{attached}$ and $M_{detached}$.}
    \label{fig:bmc}
\end{figure}
The similar behavior of the models is also expressed in the confusion matrix in figure \ref{fig:confusionmatrix}.
In this matrix we consider both the two model variations and the experiment as data-generating models.
The entry in cell $(i,j)$ describes with what probability outputs of a data-generating model $M_i$ get ascribed to model $M_j$.
Each of the generative models achieves a weight larger than $0.5$ on the diagonal, which supports that the models are distinct enough to be able to recognize themselves.
The weight for correctly recognizing the experimental results achieves a clear absolute majority in its own corresponding column, with weights of $0.24$ for each of the model variations.
This shows that, while the experimental data can be identified, the models are able to properly represent some of the processes in the experiment.
This supports the model extension as presented in section~\ref{sec:extendModel}.
The confusion matrix does not indicate a preference for $M_\mathrm{attached}$ or $M_\mathrm{detached}$.
Instead, they can be perceived as interchangeable, indicated by the high similarity in the weights associated to the two models.
We can conclude that the experimental results are well within the scope of both model variants, while the inverse does not hold true.

\section{Conclusions}
\label{sec:sum}

In the following, we provide a brief summary of our study, present our lessons learned with a particular focus on the representativeness of the experimental results, and outline perspectives for future work.

\paragraph{Summary}
This work investigated the appearance of preferential flow paths in EICP by extension of an existing numerical model and subsequent Bayesian analysis of its behaviour.
We amended an REV-scale EICP model to account for ACC as a mobile precursor calcium carbonate phase. We introduced attachment and detachment mechanisms for ACC, as well as the transformation of ACC to calcite.
The output of the model was approximated with an aPCE surrogate model to reduce the computational costs of the evaluations.
The model setting was chosen based on previous microfluidic experiments.

\paragraph{Lessons learned}
The evaluations of the extended REV-scale model confirmed the formation of preferential flow as a self-enhancing process.
In order to model preferential flow-path formation in this setting, both a mobile calcium carbonate precipitate as well as initial heterogeneity in the system were required.
By varying four unknown parameters, the number and intensity of the formed preferential flow paths could be varied, and both straight and curved flow paths appeared.
Evaluations with different inlet and outlet descriptions were shown to result in changes in the placement of the flow paths.
The influence of the unknown parameters on the numerical model was examined on parameter distributions, that were determined to produce the full spectrum of precipitation, from predominant attachment to predominant detachment.
An importance ranking of the parameters was given based on sensitivity analysis with Sobol' indices and the ranking was explained with an investigation of the expected evolution of individual spatial cells in the model domain.

The lack of prior knowledge about the initial state of ACC after precipitation was explored by comparing two variants of the model.
The comparison was performed both directly and with Bayesian multi-model comparison via a confusion matrix.
The two model variants resulted in practically the same porosity distributions for most of the examined parameter ranges, and differed only for extreme values.
This can be explained by the establishment of a dynamic equilibrium due to the interdependence of the attachment and detachment processes.
Similarly, no preference between the two model variants was observed based on a comparison with the experiment in Bayesian model comparison.
We concluded from high weights for the model variants on the experimental data, that the experimental results lie inside the scope of both model variants, while the inverse does not hold.
The certainty of the conclusions from this comparison is restricted by the scarcity of experimental data.
To fully validate the model extension, more experimental measurements are vital.

\paragraph{On the representativeness of the experimental results}
The study explored the phenomenon of preferential flow for a single,specific experimental setup and a corresponding REV-scale numerical EICP model.
Thus, the insights carry certain limitations that can be expanded on in future work.
For example, the evolving preferential flow path at the top of the microfluidic cell is likely resulting from the experimental setup considered.
This includes the placement of the in- and outlet in the upper half of the porous domain.
In contrast to field applications of biomineralization, where the porous domain is unlimited in spatial extent, the limited length and width of the microfluidic porous domain in this specific setup is expected to influence the shape and length of the resulting preferential flow path. For the occurrence or absence of a flow path, this is a biased situation.
Nevertheless, it should be noted that this setup does not limit the relevance of the physical processes involved in the self-enhancing evolution of preferential flow paths during biomineralization, which the model has been shown to replicate.
Beyond that and beyond the scope of this study, the mechanisms require further numerical analysis to assess their relevance for field applications.

\paragraph{Perspectives for further work}
Both the velocity and the critical velocity are averaged REV-scale quantities. 
Thus, $v_\mathrm{Crit}$ represents only a proxy for the true pore-scale shear and other hydrodynamic forces required to model ACC detachment on the REV scale with our phenomenological approach. 
Further pore-scale experimental investigation of ACC detachment during EICP will enable a more rigorous parameterization and modeling of ACC detachment due to more detailed process understanding of pore-scale ACC detachment. 
This could be achieved by experimental setups with in-situ pore-scale velocity measurements, e.g. by particle image velocimetry.
Finally, the very pronounced flow path observed in the experiment by \citet{weinhardt2022spatiotemporal} might be due to the experimental setup, forcing a constant flow through a quasi 2D porous media with limited size. 
Future EICP experiments could reveal the initial and boundary conditions required for preferential flow-path development by e.g. comparing porous media of varying heterogeneity and injection strategies using constant injection rates or constant injection pressure.
Ideally, such studies should be conducted in 3D experimental setups to investigate whether preferential flow path development also occurs in porous media more relevant to practical applications. 
Similar studies, although numerical and focusing on the influence of injection strategies on the amount and distribution of precipitated calcium carbonate rather than flow path formation have already been conducted \citep{Hommel2016}.

\section*{Statements and declarations}

\subsection*{Funding}
The authors would like to thank the German Research Foundation (DFG) for financial support of the project within the Cluster of Excellence ``Data-Integrated Simulation Science'' (EXC 2075 – 390740016), Collaborative Research Centre SFB 1313, Project Number 327154368 and DFG Project 432343452.
Johannes Hommel thanks the DFG for supporting this work by funding Project Number 380443677.

\subsection*{Competing interests}
The authors have no relevant financial or non-financial interests to disclose.

\subsection*{Author contributions}
Conceptualization: R. Kohlhaas, J. Hommel, F. Weinhardt, H. Class, B. Flemisch; 
Model concept: J. Hommel, F. Weinhardt, H. Class; 
Model implementation: R. Kohlhaas, J. Hommel; 
Bayesian model analysis concept: R. Kohlhaas, S. Oladyshkin; 
Bayesian model analysis implementation: R. Kohlhaas, S. Oladyshkin; 
Production and analysis of results: R. Kohlhaas, J. Hommel; 
Writing - original draft preparation: R. Kohlhaas, J. Hommel;
Writing - review and editing: F. Weinhardt, H. Class, S. Oladyshkin, B. Flemisch;
Funding acquisition: H. Class, S. Oladyshkin, B. Flemisch

\subsection*{Data availability}
The source code and datasets generated during the current study are available in the  Data Repository of the University of Stuttgart (DaRUS).
The extended EICP model can be found in~\citep{DARUS-4653_2024}.
The datasets and the source code for the surrogate models and evaluations can be found in~\citep{DARUS-4654_2024}.
These datasets will be openly accessible upon final publishing.

\begin{appendices}

\section{Surface area estimation}
The REV scale model described in section \ref{sec:extendModel} uses a relation for the surface areas, that was estimated using the experimental results of \citet{weinhardt2022spatiotemporal}, which were introduced in section \ref{sec:experiment}.

The precipitated volumes of the experiment were estimated assuming frustrum shapes of the precipitates, which has been validated based on $\mu$-XRCT scans described in the work of \citep{weinhardt2022spatiotemporal}.
The 3D reconstruction of the precipitates is described as
\begin{align}\label{eq:frustumvolume}
    V_\mathrm{frustum}=\sum_{i=1}^n \min(\mathrm{dist}(i)\tan(\alpha),H)\cdot l_{px}^2.
\end{align}
This volume was calculated using the Euclidean distance between each pixel center that is considered precipitate and the closest pixel center that is considered void space.
The angle of the frustum was approximated as $\alpha=72\deg$. 
Based on this reconstruction, the pixels were divided into empty pixels with volume $V_{px}=0$, filled pixels with volume $V_{px}=1$, and border pixels with volume $0<V_{px}<1$.
The corresponding porosities were obtained as $\phi(px) = \frac{V_{px}}{V_\mathrm{filled}}=V_{px}$.

\label{ssec:surface}
\begin{figure}[tb]\centering
	\subfigure[]{\includegraphics[width=0.08\linewidth]{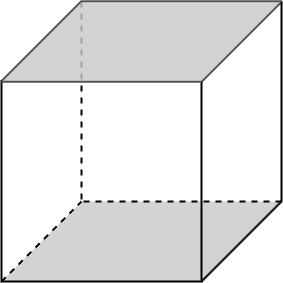}}
	\subfigure[]{\includegraphics[width=0.08\linewidth]{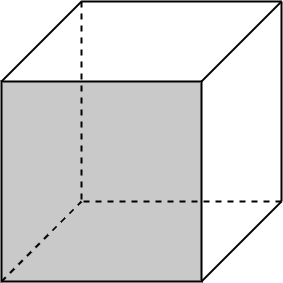}}
	\subfigure[]{\includegraphics[width=0.105\linewidth]{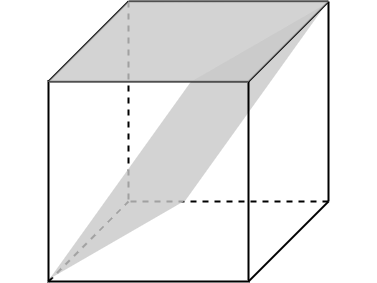}}
    \caption{Three components of the surface area estimation depending on the type of pixel, with the surface area contribution in grey: (a) fully empty, (b) completely filled and (c) border pixel.}
    \label{fig:surfaceTypes}
\end{figure}
We calculated the surface area individually for each image pixel $px$ depending its classification as void space ($\phi(px)=1$), solid matrix and solid precipitate ($\phi(px)=0$), or precipitate border ($0<\phi(px)<1$).
The surface area contributions, visualized in grey in figure \ref{fig:surfaceTypes}, are then summed to give the total surface area of each dicrete cell in the model,
\begin{align}
    A_{\mathrm{data}} = A_{\phi(px)=1}+A_{\phi(px)=0}+A_{0<\phi(px)<1}.
\end{align}

From these values we approximate the experimental relationship between the surface area and the amount of precipitate in each cell via curve fitting.
We describe the surface area $A(x,y)$ of each discrete cell $(x,y)$ in dependence of the increase in precipitate compared to the initially given calcite by introducing
\begin{align}
    \Bar{\phi} = \frac{\phi_\mathrm{precip}-\phi_\mathrm{precip,0}}{\phi_\mathrm{total,0}},
\end{align}
where $\phi_\mathrm{precip}$ is the current volume fraction of the precipitates ACC and calcite, $\phi_\mathrm{precip,0}$ the initial precipitate volume fraction for $t=0$ and $\phi_\mathrm{total,0}$ the porosity for $t=0$.

\begin{figure} [tb]\centering
	\includegraphics[width=0.65\linewidth]{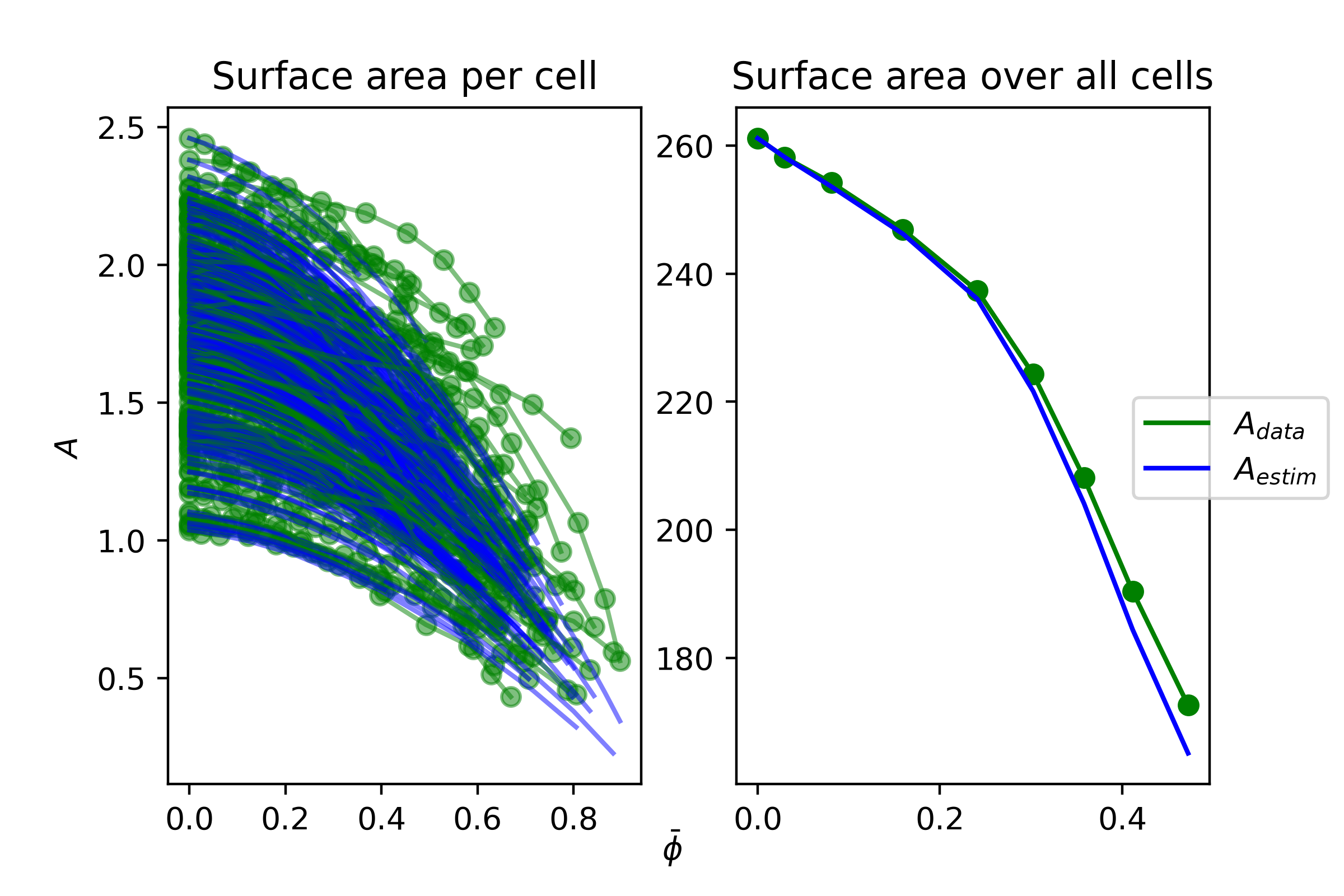}
	\caption{Estimation of the surface area to precipitate porosity ratio for each discrete cell and the full microfluidic cell.} \label{fig:surfacetoporo}
\end{figure}

We assume a parabolic relationship with the parameters $a$, $b$ and $c$,
\begin{align} \label{eq:surfacetoporo}
    A_{\mathrm{estim}}=A_0( a\cdot (\Bar{\phi})^2+b\cdot (\Bar{\phi})+c).
\end{align}
Here, $A_0$ describes the initial specific surface area at $t=0$ that was calculated from the experimental results.
We set $c=1$ to ensure the initial surface area matches the known values.
The two parameters $a$ and $b$ were calculated individually for each discrete cell from the data ${(A(x,y)_t, \Bar{\phi}(x,y)_t), t \in 0,0.5,\dots,3.5,4}$ using curve fitting and Cramer's rule.
The mean of the results was chosen as the final parameters, leading to $a= -0.75522$ and $b = -0.19683$.
The approximations that follow from this parameter choice are visualized in figure \ref{fig:surfacetoporo}.

\section{Sensitivity of the surrogate truncation scheme}\label{app:sensitivity}
The chosen truncation scheme of the aPCE-surrogate essentially defines which terms the expansion can contain, and thus which dependencies between inputs and outputs the surrogate is able to reproduce.
This influences any applied evaluation metric, but is clearly visible in sensitivity analysis.

We show an example of this effect here for hyperbolic truncation via the $q$-norm \citep{blatman2011adaptive} for $d$ input variables, maximum total degree $p$ and $q\leq1$.
For $q=1$ this approach gives the standard total-degree truncation scheme,
\begin{align}
    \mathcal{A}^{d,p,q} = \{\alpha\in\mathcal{A}^{d,p}:||\alpha||_q\leq p\},\\
    ||\alpha||_q=\left(\sum_{i=1}^d\alpha_i^q\right)^{1/q}.
\end{align}

\begin{figure} [tb]\centering
	\includegraphics[width=0.9\linewidth]{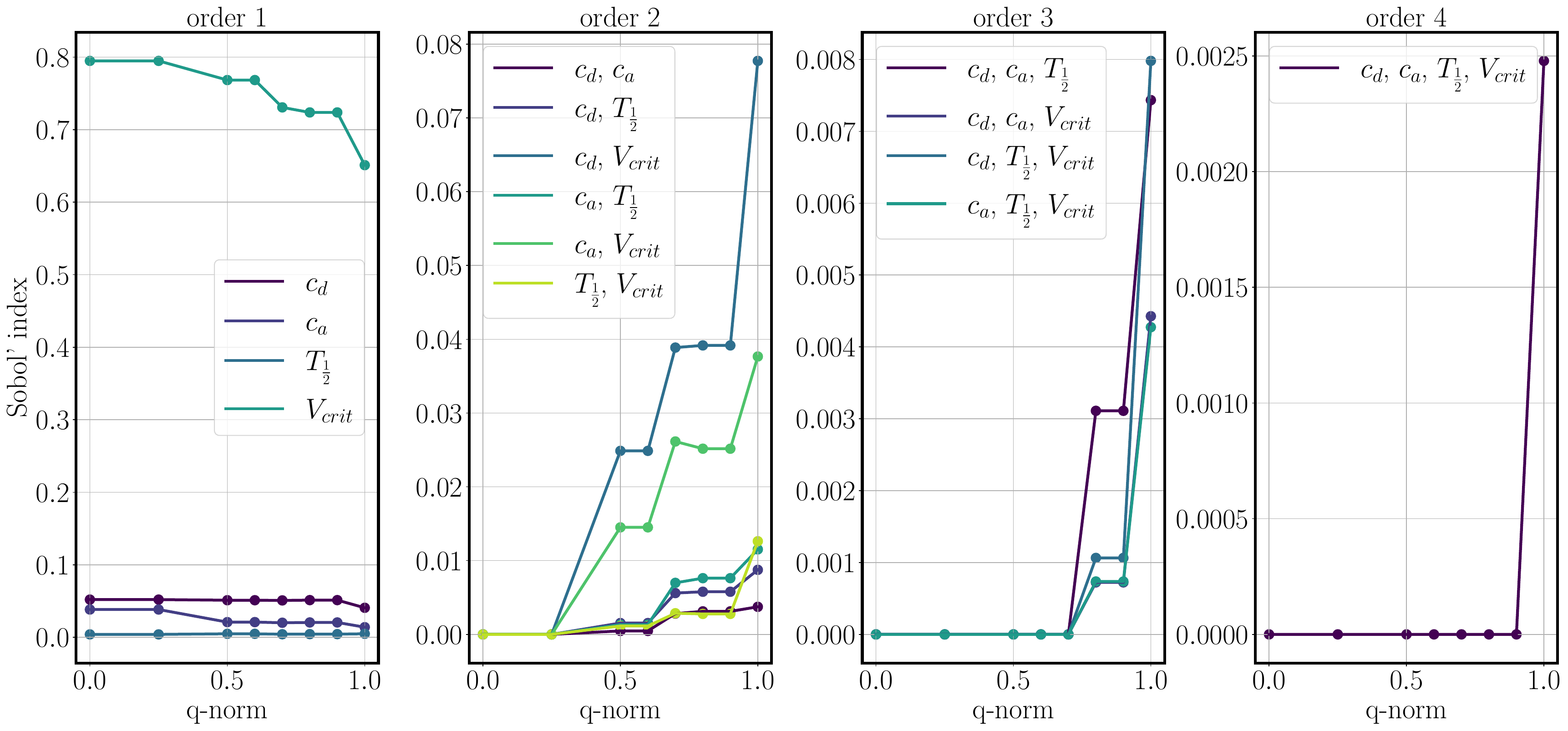}
	\caption{Sobol' indices of surrogates for model $M_{attached}$ for various $q$-norm values. Sobol' indices of different orders only activate after certain $q$ values are surpassed.} \label{fig:sobolqnorm}
\end{figure}

We compare the Sobol' indices of surrogates for model $M_\mathrm{attached}$ that were trained using different values $q$.
Figure \ref{fig:sobolqnorm} visualizes the Sobol' indices up to order four for various $q$ values.
It can clearly be seen that each order of the indices is only used once $q$ surpasses a specific value, here approximately $0.4$ for second order, $0.7$ for third order and between $0.9$ for the fourth order Sobol' indices.
This visualizes how vital the choice of $q$ is for the surrogate evaluations.
Since we want to consider as many interactions and Sobol' degrees as available in the training data, we set $q=1$ for all surrogates used in the main body of this work.




\end{appendices}


\bibliography{sn-bibliography}

\end{document}